\documentstyle[amssymb,11pt]{article}

\newlength{\minitwocolumn}
\font\teneufm=eufm10
\font\seveneufm=eufm7
\font\fiveeufm=eufm5
\newfam\eufmfam
\textfont\eufmfam=\teneufm
\scriptfont\eufmfam=\seveneufm
\scriptscriptfont\eufmfam=\fiveeufm

\makeatletter
\@addtoreset{equation}{section}
\makeatother

\newtheorem{thm}{Theorem}[section]
\newtheorem{prop}[thm]{Proposition}

\newtheorem{dfn}[thm]{Definition}
\title{\bf
\Large{\bf
ELLIPTIC DEFORMED 
SUPERALGEBRA $U_{q,p}(\widehat{{sl}}(M|N))$}
}
\begin{document}
\maketitle
\begin{center}
{TAKEO KOJIMA}
\\~\\
{\it
Department of Mathematics and Physics,
Graduate School of Science and Engineering,\\
Yamagata University, Jonan 4-3-16, Yonezawa 992-8510,
Japan\\
kojima@yz.yamagata-u.ac.jp}
\end{center}

~\\
~\\

\begin{abstract}
We introduce the elliptic superalgebra 
$U_{q,p}(\widehat{sl}(M|N))$ as one parameter deformation
of the quantum superalgebra $U_q(\widehat{sl}(M|N))$.
For an arbitrary level $k \neq 1$
we give the bosonization
of the elliptic superalgebra
$U_{q,p}(\widehat{sl}(1|2))$
and the screening currents
that commute with 
$U_{q,p}(\widehat{sl}(1|2))$
modulo total difference.
\end{abstract}

~\\
~\\

\section{Introduction}

Infinite dimensional symmetry has been an impressive
success in conformal field theory (CFT)
\cite{BPZ}.
Solvable lattice model is an off-critical
extension of CFT
and
infinite dimensional symmetry
plays an important role in algebraic analysis
of solvable lattice model \cite{JM}.
The lattice counterpart of minimal unitary CFT is
Andrews-Baxter-Forrester (ABF) model
\cite{ABF},
whose Boltzmann weights are
elliptic solutions of the Yang-Baxter equation (YBE)
of the face-type.
Among the solvable models based on YBE,
those related to elliptic solutions occupy a fundamental place.
Elliptic algebras
are certain algebraic structures
introduced to investigate these elliptic models. 
In study of $k$-fusion hierarchy of ABF model,
Konno \cite{Konno}
introduced the elliptic algebra $U_{q,p}(\widehat{sl}(2))$
and constructed
bosonization of the vertex operator
by using this algebra.
Jimbo-Konno-Odake-Shiraishi \cite{JKOS2}
constructed
the elliptic algebra $U_{q,p}(g)$ 
by dressing the usual Drinfeld currents \cite{D}
of the quantum group $U_q(g)$ for
non-twisted affine Lie algebra $g$.
In this paper
we introduce the elliptic deformed superalgebra 
$U_{q,p}(\widehat{sl}(M|N))$ as one parameter deformation
of the quantum superalgebra $U_q(\widehat{sl}(M|N))$.
We give the bosonization
of the elliptic superalgebra
$U_{q,p}(\widehat{sl}(1|2))$ and $U_{q,p}(\widehat{sl}(2|1))$
for generic level $k$, and
give the screening currents that commute with 
$U_{q,p}(\widehat{sl}(1|2))$ and 
$U_{q,p}(\widehat{sl}(2|1))$
modulo total difference.

In this paper we aim to contribute mathematical tools
for the study of super $\widehat{sl}(M|N)$-family of 
the ABF model \cite{Okado}.
Mathematical tools are
the elliptic algebra $U_{q,p}(\widehat{sl}(M|N))$
and bosonizations.
We give comments on 
$\widehat{sl}(N)$-family of the ABF model,
where such mathematical tools have been used previously 
in analogous, but simpler case.
Andrews-Baxter-Forrester \cite{ABF}
introduced the ABF model, that gives an extension of
the hard hexagon model, and derived
local height probabilities by
Baxter's corner transfer matrix method (CTM) \cite{Baxter}.
The $k$-fusion and higher-rank generalization,
that we call $\widehat{sl}(N)$-family of the ABF model,
have been studied in \cite{DJMO, DJKMO, JMO, JKMO}.
Inspired by the vertex operator approach to
the 6-vertex model \cite{JM, DFJMN, JMMN, FJMMN},
that originated from CTM,
Lukyanov-Pugai \cite{LP} studied
the vertex operator approach to the ABF model,
and derived integral representations
of multi-point local height probabilities.
In study of $k$-fusion hierarchy of ABF model,
Konno \cite{Konno}
introduced the elliptic algebra $U_{q,p}(\widehat{sl}(2))$
and constructed
bosonization of the vertex operator
by using this algebra.
The vertex operator approach to
the higher-rank generalization of the ABF model
have been studied in
\cite{AJMP, FKQ, FJMOP}.
In the vertex operator approach to
$\widehat{sl}(N)$-family of the ABF model,
bosonization of the vertex operator
played important role.
In construction of the vertex operator,
the current of the elliptic algebra 
$U_{q,p}(\widehat{sl}(N))$
and its bosonization played important roles.
In order to derive integral representation
of multi-point local height probabilities
of super $\widehat{sl}(M|N)$-family of the ABF model,
we have to construct bosonizations of the
vertex operators by using the current of the
elliptic algebra $U_{q,p}(\widehat{sl}(M|N))$,
and understand the structure of the space of state
of the model by CTM \cite{Baxter},
that has been open problem for superalgebra
$\widehat{sl}(M|N)$.

Next we give comments on 
pure mathematical aspects.
Through an attempt to understand solvable models
based on elliptic solutions of YBE,
various versions of elliptic algebras
have been introduced 
\cite{Konno, JKOS2, KK1, KK2, Sklyanin,
FIJKMY, Felder,
EF, Fronsdal1, Fronsdal2, JKOS1, ZG}.
It is important to understand
not only themselves but also relations between them.
Here we summarize some basic facts on
the elliptic quantum group ${\cal B}_{q,\lambda}(g)$
and the elliptic algebra $U_{q,p}(g)$.
The elliptic quantum group ${\cal B}_{q,\lambda}(g)$, 
was introduced by twisting the standard quantum group
$U_q(g)$ \cite{Felder, EF, Fronsdal1, Fronsdal2, JKOS1},
where $g$ is the symmetrizable Kac-Moody algebra.
The elliptic quantum group ${\cal B}_{q,\lambda}(g)$
has quasi-Hopf structure and
the elliptic algebra 
$U_{q,p}(\widehat{sl}(2))$ has
$H$-Hopf algebroid structure \cite{Konno2, Konno3}.
The realizations of the $L$-operators 
of the elliptic quantum group
${\cal B}_{q,\lambda}(g)$ were constructed in 
\cite{JKOS2, KK1, KK2} by using
the currents of the elliptic algebra $U_{q,p}(g)$ 
for $g=\widehat{sl}(N), A_2^{(2)}$.
This suggests that 
the currents of $U_{q,p}(g)$ 
give the Drinfeld currents \cite{D}
of the elliptic quantum group 
${\cal B}_{q,\lambda}(g)$.
The construction of the elliptic quantum group
${\cal B}_{q,\lambda}(g)$
has been extended to the superalgebra
$g={\widehat sl}(M|N)$ \cite{ZG}.
In this paper we introduce the elliptic algebra 
$U_{q,p}(\widehat{sl}(M|N))$.
We conjecture that the $L$-operator of
${\cal B}_{q,\lambda}(\widehat{sl}(M|N))$
is constructed by using
the currents of $U_{q,p}(\widehat{sl}(M|N))$
and that
there exists
$H$-Hopf algebroid structure for
$U_{q,p}(\widehat{sl}(M|N))$.
The bosonizations of the vertex operators
give useful information for construction of the $L$-operator,
thorough so-called Miki's construction \cite{Miki}
of the $L$-operator.
The above is background mathematical theory of 
the vertex operator.
Next we give a comment on mathematical phenomenon 
of the space of state.
Date-Jimbo-Kuniba-Miwa-Okado 
\cite{DJKMO, DJKMO2, JMO, JKMO}
found that local height probabilities of  
$\widehat{sl}(N)$-family of the ABF model
were expressed in terms of
the branching coefficients appearing in the irreducible decomposition
of the character of $\widehat{sl}(N)$ \cite{Kac-Peterson, GKO}.
In order to extend this to 
super $\widehat{sl}(M|N)$-family of the ABF model,
we have to know the character formulae of
the superalgebra $\widehat{sl}(M|N)$,
that gives the affine generalization of formulae
\cite{Serganova, Jonathan}.

The text is organized as follows.
In section 2, after preparing the notations and
giving the definition of 
the quantum group $U_q(\widehat{sl}(M|N))$,
we introduce the elliptic deformed superalgebra
$U_{q,p}(\widehat{sl}(M|N))$.
Our approach is based on the dressing procedure
of the Drinfeld current
of the quantum group.
In section 3 
we give bosonizations of
the superalgebra
$U_{q}(g), U_{q,p}(g)$
$(g=\widehat{sl}(1|2), \widehat{sl}(2|1))$
for an arbitrary level $k$.
We give
the screening currents
that commute with 
$U_{q}(g), U_{q,p}(g)$ 
$(g=\widehat{sl}(1|2), \widehat{sl}(2|1))$
modulo total difference.
In appendix we summarize some useful formulae of
bosonizations and screening currents.

\section{Elliptic deformed superalgebra 
$U_{q,p}(\widehat{sl}(M|N))$}

In this section we introduce the elliptic
superalgebra $U_{q,p}(\widehat{sl}(M|N))$.
Kac \cite{Kac} 
introduced the superalgebra 
generalization of contragredient Lie algebra.
Van de Leur \cite{Leur} classified 
the contragredient superalgebra $g$
of finite growth.
Yamane \cite{Yamane} introduced
quantum affine superalgebra $U_q(g)$ and
constructed the Drinfeld currents.
We give
elliptic deformation of the
quantum affine superalgebra by developing
the dressing procedure 
\cite{JKOS2}.

\subsection{Quantum superalgebra $U_q(\widehat{sl}(M|N))$}

In this section we review
the Drinfeld realization
of the quantum superalgebra $U_q(\widehat{sl}(M|N))$
for $M,N=1,2,3,\cdots$ \cite{Yamane}.
We restrict our consideration to $M \neq N$.
The quantum superalgebra $U_q(\widehat{sl}(M|N))$
in \cite{Yamane} is a $q$-deformation of the universal
enveloping algebra of $\widehat{sl}(M|N)$
\cite{Leur}.
Hereafter we fix a complex number $q \neq 0, |q|<1$.
Let us set
\begin{eqnarray}
[x,y]=xy-yx,~~\{x,y\}=xy+yx,~~[a]_q=\frac{q^a-q^{-a}}{
q-q^{-1}}.
\end{eqnarray}
The Cartan matrix of the Lie superalgebra
$\widehat{sl}(M|N)$ is given by
\begin{eqnarray}
(A_{i,j})_{0\leqq i,j \leqq M+N-1}=\left(\begin{array}{ccccccccccc}
0&-1&0&\cdots & & & & &\cdots &0&1\\
-1&2&-1&\cdots & & & & &\cdots &0&0\\
0&-1&2&\cdots & & & & &\cdots &\cdots &\cdots \\
\cdots&\cdots&\cdots&\cdots &-1&\cdots & & & & & \\
 & &\cdots&-1&2&-1&\cdots& & & & \\
 & & &\cdots &-1&0&1&\cdots& & & \\
 & & & & &\cdots 1&-2&1&\cdots & & \\
 & & & & &\cdots &1 &\cdots&\cdots&\cdots &\cdots\\
\cdots &\cdots &\cdots & & & & &\cdots&-2& 1&0\\
0&0&\cdots & & & & &\cdots&1&-2&1\\
1&0&\cdots & & & & &\cdots &0&1&-2 
\end{array}\right),
\end{eqnarray}
where the diagonal part is
$(A_{i,i})_{0\leqq i \leqq M+N-1}
=(0,\overbrace{2, \cdots, 2}^{M-1},0,
\overbrace{-2, \cdots, -2}^{N-1})$.

\begin{dfn}~~\cite{Yamane}~
The generators of
the quantum superalgebra $U_q(\widehat{sl}(M|N))$,
which we call the Drinfeld generators,
are given by
\begin{eqnarray}
x_{i,m}^\pm,~a_{i,n},~h_i,~c,~~(1\leqq i \leqq M+N-1,
m \in {\mathbb Z}, n \in {\mathbb Z}_{\neq 0}).
\end{eqnarray}
Defining relations are
\begin{eqnarray}
&&~c : {\rm central},~[h_i,a_{j,m}]=0,\\
&&~[a_{i,m},a_{j,n}]=\frac{[A_{i,j}m]_q[cm]_q}{m}q^{-c|m|}
\delta_{m+n,0},\\
&&~[h_i,x_j^\pm(z)]=\pm A_{i,j}x_j^\pm(z),\\
&&~[a_{i,m}, x_j^+(z)]=\frac{[A_{i,j}m]_q}{m}
q^{-c|m|} z^m x_j^+(z),\\
&&~[a_{i,m}, x_j^-(z)]=-\frac{[A_{i,j}m]_q}{m}
z^m x_j^-(z),\\
&&(z_1-q^{\pm A_{i,j}}z_2)
x_i^\pm(z_1)x_j^\pm(z_2)
=
(q^{\pm A_{j,i}}z_1-z_2)
x_j^\pm(z_2)x_i^\pm(z_1),~~{\rm for}~|A_{i,j}|\neq 0,
\\
&&
x_i^\pm(z_1)x_j^\pm(z_2)
=
x_j^\pm(z_2)x_i^\pm(z_1),~~{\rm for}~|A_{i,j}|=0, (i,j)\neq (M,M),
\\
&&
\{x_M^\pm(z_1), x_M^\pm(z_2)\}=0,\\
&&~[x_i^+(z_1),x_j^-(z_2)]
=\frac{\delta_{i,j}}{(q-q^{-1})z_1z_2}
\left(
\delta(q^{-c}z_1/z_2)\psi_i^+(q^{\frac{c}{2}}z_2)-
\delta(q^{c}z_1/z_2)\psi_i^-(q^{-\frac{c}{2}}z_2)
\right), \nonumber\\
&& ~~~~~{\rm for}~~(i,j) \neq (M,M),\\
&&~\{x_M^+(z_1),x_M^-(z_2)\}
=\frac{1}{(q-q^{-1})z_1z_2}
\left(
\delta(q^{-c}z_1/z_2)\psi_M^+(q^{\frac{c}{2}}z_2)-
\delta(q^{c}z_1/z_2)\psi_M^-(q^{-\frac{c}{2}}z_2)
\right), \nonumber\\
\\
&& 
\left(
x_i^\pm(z_{1})
x_i^\pm(z_{2})
x_j^\pm(z)-(q+q^{-1})
x_i^\pm(z_{1})
x_j^\pm(z)
x_i^\pm(z_{2})
+x_j^\pm(z)
x_i^\pm(z_{1})
x_i^\pm(z_{2})\right)\nonumber\\
&&+\left(z_1 \leftrightarrow z_2\right)=0,
~~~{\rm for}~|A_{i,j}|=1,~i\neq M,\\
\nonumber
\\
&&\left(
x_M^\pm(z_1)x_{M+1}^\pm(w_1)x_M^\pm(z_2)x_{M-1}^\pm(w_2)
-q^{-1}x_M^\pm(z_1)x_{M+1}^\pm(w_1)x_{M-1}^\pm(w_2)x_M^\pm(z_2)
\right.\nonumber\\
&&-q
x_M^\pm(z_1)
x_M^\pm(z_2)
x_{M-1}^\pm(w_2)
x_{M+1}^\pm(w_1)
+x_M^\pm(z_1)
x_{M-1}^\pm(w_2)
x_M^\pm(z_2)
x_{M+1}^\pm(w_1)\nonumber\\
&&+x_{M+1}^\pm(w_1)
x_{M}^\pm(z_2)
x_{M-1}^\pm(w_2)
x_{M}^\pm(z_1)
-q^{-1}
x_{M+1}^\pm(w_1)
x_{M-1}^\pm(w_2)
x_{M}^\pm(z_2)x_{M}^\pm(z_1)
\nonumber\\
&&
\left.
-q
x_{M}^\pm(z_2)
x_{M-1}^\pm(w_2)
x_{M+1}^\pm(w_1)
x_{M}^\pm(z_1)
+
x_{M-1}^\pm(w_2)
x_{M}^\pm(z_2)
x_{M+1}^\pm(w_1)
x_{M}^\pm(z_1)
\right)\nonumber\\
&&+(z_1 \leftrightarrow z_2)=0,
\end{eqnarray}
where we have used
$\delta(z)=\sum_{m \in {\mathbb Z}}z^m$.
Here we have set the generating functions
\begin{eqnarray}
x_j^\pm(z)&=&
\sum_{m \in {\mathbb Z}}x_{j,m}^\pm z^{-m-1},\\
\psi_i^+(q^{\frac{c}{2}}z)&=&q^{h_i}
\exp\left(
(q-q^{-1})\sum_{m>0}a_{i,m}z^{-m}
\right),\\
\psi_i^-(q^{-\frac{c}{2}}z)&=&q^{-h_i}
\exp\left(-(q-q^{-1})\sum_{m>0}a_{i,-m}z^m\right).
\end{eqnarray}
\end{dfn}
We
changed the gauge of boson $a_{i,m}$
from those of \cite{Yamane}.
In what follows we assume $c \in {\mathbb C}$.

\subsection{Elliptic deformed superalgebra
$U_{q,p}(\widehat{sl}(M|N))$}

In this section we introduce the elliptic
superalgebra $U_{q,p}(\widehat{sl}(M|N))$ for $M,N=1,2,3,\cdots, (M \neq N)$.
Let us introduce a deformation parameter $r$ such that
\begin{eqnarray}
r,~ r^*=r-c>0.
\end{eqnarray}
We often use the parameterization.
\begin{eqnarray}
&&p=q^{2r}=e^{-\frac{2\pi i}{\tau}},
~p^*=q^{2r^*}=e^{-\frac{2\pi i}{\tau^*}},
~z=q^{2u},~w=q^{2v}.
\end{eqnarray}
We have $r\tau=r^*\tau^*$.
Let us set the Jacobi theta functions
$[u]$, $[u]^*$ by
\begin{eqnarray}
[u]=q^{\frac{u^2}{r}-u}\frac{\Theta_p(q^{2u})}
{(p;p)_\infty^3},~~~
[u]^*=q^{\frac{u^2}{r^*}-u}\frac{\Theta_{p^*}(q^{2u})}
{(p^*;p^*)_\infty^3}.
\end{eqnarray}
Here we have used the standard symbols.
\begin{eqnarray}
&&\Theta_p(z)=(p;p)_\infty (z;p)_\infty (pz^{-1};p)_\infty,\\
&&(z;t_1,\cdots,t_k)_\infty=\prod_{n_1,\cdots,n_k \geqq 0}
(1-zt_1^{n_1}\cdots t_k^{n_k}).
\end{eqnarray}

\begin{dfn}~~~
The elliptic superalgebra 
$U_{q,p}(\widehat{sl}(M|N))$ is generated by
the currents (operator valued function) and elements
\begin{eqnarray}
E_j(z),~F_j(z),~
B_{j,n},~~h_j,~~c~~
(1\leqq j \leqq M+N-1,
n \in {\mathbb Z}_{\neq 0}).
\end{eqnarray}
The defining relations are
given as follows.\\
For $1\leqq i,j \leqq M+N-1$,
the relations are
\begin{eqnarray}
&&~c : {\rm central},~[h_i,B_{j,m}]=0,
\label{def:elliptic1}\\
&&~[B_{i,m},B_{j,n}]=\frac{[A_{i,j}m]_q[cm]_q}{m}
\frac{[r^*m]_q}{[rm]_q}
\delta_{m+n,0},
\label{def:elliptic2}
\\
&&~[h_i,E_j(z)]=A_{i,j}E_j(z),~[h_i,F_j(z)]=-A_{i,j}F_j(z),
\label{def:elliptic3}
\\
&&~[B_{i,m}, E_j(z)]=\frac{[A_{i,j}m]_q}{m}
z^m E_j(z),
~[B_{i,m}, F_j(z)]=-\frac{[A_{i,j}m]_q}{m}\frac{[r^*m]_q}{
[rm]_q}
z^m F_j(z).
\label{def:elliptic4}
\end{eqnarray}
For $1\leqq i,j \leqq M+N-1$ such that $(i,j)\neq (M,M)$,
the relations are
\begin{eqnarray}
&&~
\left[u_1-u_2-\frac{A_{i,j}}{2}\right]^*
E_i(z_1)E_j(z_2)
=
\left[u_1-u_2+\frac{A_{i,j}}{2}\right]^*
E_j(z_2)E_i(z_1),
\label{def:elliptic5}
\\
&&~\left[u_1-u_2+\frac{A_{i,j}}{2}\right]
F_i(z_1)F_j(z_2)
=
\left[u_1-u_2-\frac{A_{i,j}}{2}\right]
F_j(z_2)F_i(z_1),
\label{def:elliptic6}
\\
&&~[E_i(z_1),F_j(z_2)]
=\frac{\delta_{i,j}}{(q-q^{-1})z_1z_2}
\left(
\delta(q^{-c}z_1/z_2)H_i(q^{r}z_2)-
\delta(q^{c}z_1/z_2)H_i(q^{-r}z_2)
\right),
\label{def:elliptic7}
\\\nonumber\\
&&\{E_M(z_1),E_M(z_2)\}=0,~~~
\{F_M(z_1),F_M(z_2)\}=0,
\label{def:elliptic8}
\\
&&\{E_M(z_1),F_M(z_2)\}
=\frac{1}{(q-q^{-1})z_1z_2}
\left(
\delta(q^{-c}z_1/z_2)H_M(q^{r}z_2)-
\delta(q^{c}z_1/z_2)H_M(q^{-r}z_2)
\right).\nonumber\\
\label{def:elliptic9}
\end{eqnarray}
For $1\leqq i,j \leqq M+N-1$, the relations are
\begin{eqnarray}
&&H_i(z_1)H_j(z_2)=\frac{
[u_2-u_1-\frac{A_{i,j}}{2}]^*
[u_2-u_1+\frac{A_{i,j}}{2}]}{
[u_2-u_1+\frac{A_{i,j}}{2}]^*
[u_2-u_1-\frac{A_{i,j}}{2}]}
H_j(z_2)H_i(z_1),
\label{def:elliptic10}
\\
&&H_i(z_1)E_j(z_2)=
\frac{
[u_1-u_2+\frac{r^*}{2}+\frac{A_{i,j}}{2}]^*}{
[u_1-u_2+\frac{r^*}{2}-\frac{A_{i,j}}{2}]^*}
E_j(z_2)H_i(z_1),
\label{def:elliptic11}
\\
&&H_i(z_1)F_j(z_2)=
\frac{
[u_1-u_2+\frac{r}{2}+\frac{A_{i,j}}{2}]}{
[u_1-u_2+\frac{r}{2}-\frac{A_{i,j}}{2}]}
F_j(z_2)H_i(z_1).
\label{def:elliptic12}
\end{eqnarray}
For $1\leqq i,j \leqq M+N-1$ $(i\neq M)$ 
such that $|A_{i,j}|=1$,
they satisfy
the Serre relations
\begin{eqnarray}
&&
~~\left(
E_i(z_1)E_i(z_2)E_j(z)
\frac{
\{q^{A_{i,j}}\frac{z}{z_1}\}^* 
\{q^{A_{i,j}}\frac{z}{z_2}\}^*}{
\{q^{-A_{i,j}}\frac{z}{z_1}\}^*
\{q^{-A_{i,j}}\frac{z}{z_2}\}^*}
\left(\frac{z}{z_2}\right)^{\frac{1}{r^*}A_{i,j}}
\right.\nonumber\\
&&-(q+q^{-1})
E_i(z_1)E_j(z)E_i(z_2)
\frac{
\{q^{A_{i,j}}\frac{z}{z_1}\}^* 
\{q^{A_{i,j}}\frac{z_2}{z}\}^*}{
\{q^{-A_{i,j}}\frac{z}{z_1}\}^* 
\{q^{-A_{i,j}}\frac{z_2}{z}\}^*}
\nonumber
\\
&&\left.
+E_j(z)E_i(z_1)E_i(z_2)
\frac{
\{q^{A_{i,j}}\frac{z_1}{z}\}^* 
\{q^{A_{i,j}}\frac{z_2}{z}\}^*}{
\{q^{-A_{i,j}}\frac{z_1}{z}\}^* 
\{q^{-A_{i,j}}\frac{z_2}{z}\}^*}
\left(\frac{z_1}{z}\right)^{\frac{1}{r^*}A_{i,j}}
\right)
\frac{\{q^{A_{i,i}}\frac{z_2}{z_1}\}^*}{
\{q^{-A_{i,i}}\frac{z_2}{z_1}\}^*}
z_1^{-\frac{1}{r^*}(A_{i,i}+A_{i,j})}\nonumber\\
&&+(z_1 \leftrightarrow z_2)
=0,\label{eqn:Serre1}
\\
&&~~\left(
F_i(z_1)F_i(z_2)F_j(z)
\frac{
\{q^{-A_{i,j}}\frac{z}{z_1}\} 
\{q^{-A_{i,j}}\frac{z}{z_2}\}}{
\{q^{A_{i,j}}\frac{z}{z_1}\} 
\{q^{A_{i,j}}\frac{z}{z_2}\}}
\left(\frac{z_2}{z}\right)^{\frac{1}{r}}
\right.\nonumber\\
&&-(q+q^{-1})
F_i(z_1)F_j(z)F_i(z_2)
\frac{
\{q^{-A_{i,j}}\frac{z}{z_1}\} 
\{q^{-A_{i,j}}\frac{z_2}{z}\}}{
\{q^{A_{i,j}}\frac{z}{z_1}\} 
\{q^{A_{i,j}}\frac{z_2}{z}\}}
\nonumber
\\
&&\left.
+
F_j(z)F_i(z_1)F_i(z_2)
\frac{
\{q^{-A_{i,j}}\frac{z_1}{z}\} 
\{q^{-A_{i,j}}\frac{z_2}{z}\}}{
\{q^{A_{i,j}}\frac{z_1}{z}\} 
\{q^{A_{i,j}}\frac{z_2}{z}\}}
\left(\frac{z}{z_1}\right)^{\frac{1}{r}}
\right)
\frac{\{q^{-A_{i,i}}\frac{z_2}{z_1}\}}{
\{q^{A_{i,i}}\frac{z_2}{z_1}\}}
z_1^{\frac{1}{r}(A_{i,i}+A_{i,j})}
\nonumber\\
&&+(z_1 \leftrightarrow z_2)=0,
\label{eqn:Serre2}
\end{eqnarray}
and
\begin{eqnarray}
&&
\left(
E_M(z_1)E_{M+1}(w_1)E_M(z_2)E_{M-1}(w_2)
\frac{
\{\frac{qw_1}{z_1}\}^*
\{\frac{qz_2}{w_1}\}^*
\{\frac{w_2}{qz_1}\}^*
\{\frac{w_2}{qz_2}\}^*}
{
\{\frac{w_1}{qz_1}\}^*
\{\frac{z_2}{qw_1}\}^*
\{\frac{qw_2}{z_1}\}^*
\{\frac{qw_2}{z_2}\}^*
}
\left(\frac{w_2}{z_2}\right)^{\frac{1}{r^*}}
\right.\nonumber\\
&&
-q^{-1}
E_M(z_1)E_{M+1}(w_1)E_{M-1}(w_2)E_M(z_2)
\frac{
\{\frac{qw_1}{z_1}\}^*
\{\frac{w_2}{qz_1}\}^*
\{\frac{qz_2}{w_1}\}^*
\{\frac{z_2}{qw_2}\}^*}
{
\{\frac{w_1}{qz_1}\}^*
\{\frac{qw_2}{z_1}\}^*
\{\frac{z_2}{qw_1}\}^*
\{\frac{qz_2}{w_2}\}^*
}
\nonumber\\
&&-q
E_M(z_1)
E_M(z_2)
E_{M-1}(w_2)
E_{M+1}(w_1)
\frac{
\{\frac{w_2}{qz_1}\}^*
\{\frac{w_2}{qz_2}\}^*
\{\frac{qw_1}{z_1}\}^*
\{\frac{qw_1}{z_2}\}^*}
{
\{\frac{qw_2}{z_2}\}^*
\{\frac{qw_1}{z_1}\}^*
\{\frac{w_1}{qz_2}\}^*
\{\frac{z_2}{qw_2}\}^*
}
\left(\frac{w_2}{w_1}\right)^{\frac{1}{r^*}}
\nonumber\\
&&
+
E_M(z_1)
E_{M-1}(w_2)
E_M(z_2)
E_{M+1}(w_1)
\frac{
\{\frac{w_2}{qz_1}\}^*
\{\frac{z_2}{qz_1}\}^*
\{\frac{qw_1}{z_1}\}^*
\{\frac{qw_1}{z_2}\}^*}
{
\{\frac{qw_2}{z_1}\}^*
\{\frac{qz_2}{z_1}\}^*
\{\frac{w_1}{qz_1}\}^*
\{\frac{w_1}{qz_2}\}^*
}
\left(\frac{z_2}{w_1}\right)^{\frac{1}{r^*}}
\nonumber\\
&&
+
E_{M+1}(w_1)
E_{M}(z_2)
E_{M-1}(w_2)
E_{M}(z_1)
\frac{
\{\frac{qz_2}{w_1}\}^*
\{\frac{w_2}{qz_2}\}^*
\{\frac{qz_1}{w_1}\}^*
\{\frac{z_1}{qw_2}\}^*}
{
\{\frac{z_2}{qw_1}\}^*
\{\frac{qw_2}{z_2}\}^*
\{\frac{z_1}{qw_1}\}^*
\{\frac{qz_1}{w_2}\}^*
}
\left(\frac{w_1}{z_2}\right)^{\frac{1}{r^*}}
\nonumber\\
&&-q^{-1}
E_{M+1}(w_1)
E_{M-1}(w_2)
E_{M}(z_2)
E_{M}(z_1)
\frac{
\{\frac{qz_2}{w_1}\}^*
\{\frac{z_2}{qw_2}\}^*
\{\frac{qz_1}{w_1}\}^*
\{\frac{z_1}{qw_2}\}^*}
{
\{\frac{z_2}{qw_1}\}^*
\{\frac{qz_2}{w_2}\}^*
\{\frac{z_1}{qw_1}\}^*
\{\frac{qz_1}{w_2}\}^*
}\left(\frac{w_1}{w_2}\right)^{\frac{1}{r^*}}
\nonumber\\
&&
-q
E_{M}(z_2)
E_{M-1}(w_2)
E_{M+1}(w_1)
E_{M}(z_1)
\frac{
\{\frac{w_2}{qz_2}\}^*
\{\frac{w_1}{qz_2}\}^*
\{\frac{qz_1}{w_2}\}^*
\{\frac{qz_1}{w_1}\}^*}
{
\{\frac{qw_2}{z_2}\}^*
\{\frac{qw_1}{z_2}\}^*
\{\frac{z_1}{qw_2}\}^*
\{\frac{z_1}{qw_1}\}^*}
\nonumber\\
&&
+\left.
E_{M-1}(w_2)
E_{M}(z_2)
E_{M+1}(w_1)
E_{M}(z_1)
\frac{
\{\frac{z_2}{qw_2}\}^*
\{\frac{qw_1}{z_1}\}^*
\{\frac{z_1}{qw_2}\}^*
\{\frac{qz_1}{w_1}\}^*}
{
\{\frac{qz_2}{w_2}\}^*
\{\frac{w_1}{qz_1}\}^*
\{\frac{qz_1}{w_2}\}^*
\{\frac{z_1}{qw_1}\}^*
}\left(\frac{z_2}{w_2}\right)^{\frac{1}{r^*}}
\right)\nonumber\\
&&+(z_1 \leftrightarrow z_2)=0,\label{eqn:Serre3}
\nonumber\\
\\
&&
\left(
F_M(z_1)F_{M+1}(w_1)F_M(z_2)F_{M-1}(w_2)
\frac{
\{\frac{w_1}{qz_1}\}
\{\frac{z_2}{qw_1}\}
\{\frac{qw_2}{z_1}\}
\{\frac{qw_2}{z_2}\}}
{
\{\frac{qw_1}{z_1}\}
\{\frac{qz_2}{w_1}\}
\{\frac{w_2}{qz_1}\}
\{\frac{w_2}{qz_2}\}
}
\left(\frac{z_2}{w_2}\right)^{\frac{1}{r}}
\right.\nonumber\\
&&
-q^{-1}
F_M(z_1)F_{M+1}(w_1)F_{M-1}(w_2)F_M(z_2)
\frac{
\{\frac{w_1}{qz_1}\}
\{\frac{qw_2}{z_1}\}
\{\frac{z_2}{qw_1}\}
\{\frac{qz_2}{w_2}\}}
{
\{\frac{qw_1}{z_1}\}
\{\frac{w_2}{qz_1}\}
\{\frac{qz_2}{w_1}\}
\{\frac{z_2}{qw_2}\}
}\nonumber\\
&&-q
F_M(z_1)
F_M(z_2)
F_{M-1}(w_2)
F_{M+1}(w_1)
\frac{
\{\frac{qw_2}{z_1}\}
\{\frac{qw_2}{z_2}\}
\{\frac{w_1}{qz_1}\}
\{\frac{w_1}{qz_2}\}}
{
\{\frac{w_2}{qz_2}\}
\{\frac{w_1}{qz_1}\}
\{\frac{qw_1}{z_2}\}
\{\frac{qz_2}{w_2}\}
}
\left(\frac{w_1}{w_2}\right)^{\frac{1}{r}}
\nonumber\\
&&
+
F_M(z_1)
F_{M-1}(w_2)
F_M(z_2)
F_{M+1}(w_1)
\frac{
\{\frac{qw_2}{z_1}\}
\{\frac{qz_2}{z_1}\}
\{\frac{w_1}{qz_1}\}
\{\frac{w_1}{qz_2}\}}
{
\{\frac{w_2}{qz_1}\}
\{\frac{z_2}{qz_1}\}
\{\frac{qw_1}{z_1}\}
\{\frac{qw_1}{z_2}\}
}
\left(\frac{w_1}{z_2}\right)^{\frac{1}{r}}
\nonumber\\
&&
+F_{M+1}(w_1)
F_{M}(z_2)
F_{M-1}(w_2)
F_{M}(z_1)
\frac{
\{\frac{z_2}{qw_1}\}
\{\frac{qw_2}{z_2}\}
\{\frac{z_1}{qw_1}\}
\{\frac{qz_1}{w_2}\}}
{
\{\frac{qz_2}{w_1}\}
\{\frac{w_2}{qz_2}\}
\{\frac{qz_1}{w_1}\}
\{\frac{z_1}{qw_2}\}
}\left(\frac{z_2}{w_1}\right)^{\frac{1}{r}}
\nonumber\\
&&-q^{-1}
F_{M+1}(w_1)
F_{M-1}(w_2)
F_{M}(z_2)
F_{M}(z_1)
\frac{
\{\frac{z_2}{qw_1}\}
\{\frac{qz_2}{w_2}\}
\{\frac{z_1}{qw_1}\}
\{\frac{qz_1}{w_2}\}}
{
\{\frac{qz_2}{w_1}\}
\{\frac{z_2}{qw_2}\}
\{\frac{qz_1}{w_1}\}
\{\frac{z_1}{qw_2}\}
}\left(\frac{w_2}{w_1}\right)^{\frac{1}{r}}
\nonumber\\
&&
-q
F_{M}(z_2)
F_{M-1}(w_2)
F_{M+1}(w_1)
F_{M}(z_1)
\frac{
\{\frac{qw_2}{z_2}\}
\{\frac{qw_1}{z_2}\}
\{\frac{z_1}{qw_2}\}
\{\frac{z_1}{qw_1}\}}
{
\{\frac{w_2}{qz_2}\}
\{\frac{w_1}{qz_2}\}
\{\frac{qz_1}{w_2}\}
\{\frac{qz_1}{w_1}\}}
\nonumber\\
&&
+\left.
F_{M-1}(w_2)
F_{M}(z_2)
F_{M+1}(w_1)
F_{M}(z_1)
\frac{
\{\frac{qz_2}{w_2}\}
\{\frac{w_1}{qz_1}\}
\{\frac{qz_1}{w_2}\}
\{\frac{z_1}{qw_1}\}}
{
\{\frac{z_2}{qw_2}\}
\{\frac{qw_1}{z_1}\}
\{\frac{z_1}{qw_2}\}
\{\frac{qz_1}{w_1}\}
}\left(\frac{w_2}{z_2}\right)^{\frac{1}{r}}
\right)\nonumber\\
&&+(z_1 \leftrightarrow z_2)=0.
\label{eqn:Serre4}
\end{eqnarray}
Here we have used the abbreviations
\begin{eqnarray}
\{z\}^*=(p^*z;p^*)_\infty,~~
\{z\}=(pz;p)_\infty.\label{def:abb}
\end{eqnarray}
\end{dfn}

\subsection{Dressing construction}

In this section we construct
$U_{q,p}(\widehat{sl}(M|N))$ from
$U_q(\widehat{sl}(M|N))$ by developing the dressing procedure
\cite{JKOS2}.

\begin{dfn}~~~
Let us introduce the dressing operators
$u_j^\pm(z,p),~(1\leqq j \leqq M+N-1)$ by
\begin{eqnarray}
&&u_j^+(z,p)=\exp\left(\sum_{m>0}
\frac{1}{[r^*m]_q}a_{j,-m}(q^rz)^m\right),\\
&&u_j^-(z,p)=\exp\left(-\sum_{m>0}
\frac{1}{[rm]_q}a_{j,m}(q^{-r}z)^{-m}\right).
\end{eqnarray}
\end{dfn}

Straightforward calculations show the following propositions.

\begin{prop}~~~For $1\leqq i,j \leqq M+N-1$,
we have
\begin{eqnarray}
&&u_i^+(z_1,p)x_j^+(z_2)=\frac{
(p^*q^{A_{i,j}}z_1/z_2:p^*)_\infty}{
(p^*q^{-A_{i,j}}z_1/z_2;p^*)_\infty}
x_j^+(z_2)u_i^+(z_1,p),\\
&&u_i^+(z_1,p)x_j^-(z_2)=
\frac{
(p^*q^{-A_{i,j}+c}z_1/z_2:p^*)_\infty}{
(p^*q^{A_{i,j}+c}z_1/z_2;p^*)_\infty}
x_j^-(z_2)u_i^+(z_1,p),\\
&&u_i^-(z_1,p)x_j^+(z_2)=
\frac{
(pq^{-A_{i,j}-c}z_1/z_2:p)_\infty}{
(pq^{A_{i,j}-c}z_1/z_2;p)_\infty}
x_j^+(z_2)u_i^-(z_1,p),\\
&&u_i^-(z_1,p)x_j^-(z_2)=
\frac{
(pq^{A_{i,j}}z_1/z_2:p)_\infty}{
(pq^{-A_{i,j}}z_1/z_2;p)_\infty}
x_j^-(z_2)u_i^-(z_1,p),\\
&&u_i^+(z_1,p)u_j^-(z_2,p)=\frac{
(pq^{-A_{i,j}-c}z_1/z_2;p)_\infty 
(p^*q^{A_{i,j}+c}z_1/z_2;p^*)_\infty}{
(pq^{A_{i,j}-c}z_1/z_2;p)_\infty 
(p^*q^{-A_{i,j}+c}z_1/z_2;p^*)_\infty}u_j^-(z_2,p)u_i^+(z_1,p).\nonumber\\
\end{eqnarray}
\end{prop}

\begin{dfn}~~~
We define the dressing currents $e_j(z,p), f_j(z,p),
\psi_j^\pm(z,p)$, $(1\leqq j \leqq M+N-1)$ by
\begin{eqnarray}
&&e_j(z,p)=u_j^+(z,p)x_j^+(z),\\
&&f_j(z,p)=x_j^-(z)u_j^-(z,p),\\
&&\psi_j^+(z,p)=
u_j^+(q^{\frac{c}{2}}z,p)\psi_j^+(z)u_j^-(q^{-\frac{c}{2}}z,p),\\
&&\psi_j^-(z,p)=
u_j^+(q^{-\frac{c}{2}}z,p)\psi_j^-(z)u_j^-(q^{\frac{c}{2}}z,p).
\end{eqnarray}
\end{dfn}

\begin{prop}~~~
The currents $e_i(z,p), f_i(z,p)$ and $a_{i,n}, h_i, c$,
$(1\leqq i \leqq M+N-1, 
n \in {\mathbb Z}_{\neq 0})$
satisfy the following relations
\begin{eqnarray}
&&~c : {\rm central},~[h_i,a_{j,m}]=0,\\
&&~[a_{i,m},a_{j,n}]=\frac{[A_{i,j}m]_q[cm]_q}{m}q^{-c|m|}
\delta_{m+n,0},\\
&&~[h_i,e_j(z,p)]=A_{i,j}e_j(z,p),
~~[h_i,f_j(z,p)]=-A_{i,j}f_j(z,p),\\
&&~[a_{i,m}, e_j(z,p)]=\frac{[A_{i,j}m]_q}{m}z^me_j(z,p)\times
\left\{
\begin{array}{cc}
\frac{[rm]_q}{[r^*m]_q}, &~~(m>0)\\
q^{cm}, &~~(m<0)
\end{array}
\right.,\\
&&~[a_{i,m}, f_j(z,p)]=
-\frac{[A_{i,j}m]_q}{m}z^m f_j(z,p)\times
\left\{
\begin{array}{cc}
1,&~~(m>0)\\
\frac{[r^*m]_q}{[rm]_q}q^{cm}, &~~(m<0)
\end{array}
\right.,\\
&&z_1 \Theta_{p^*}(q^{A_{i,j}}z_2/z_1)
e_i(z_1,p)e_j(z_2,p)\nonumber\\
&&=-z_2 \Theta_{p^*}(q^{A_{j,i}}z_2/z_1) 
e_j(z_2,p)e_i(z_1,p),~~{\rm for}~|A_{i,j}|\neq 0,\\
&&
~[e_i(z_1,p),e_j(z_2,p)]=0,~~{\rm for}~|A_{i,j}|=0, 
(i,j)\neq (M,M),\\
&&
\{e_M(z_1,p),e_M(z_2,p)\}=0,
\\
&&z_1 \Theta_{p}(q^{-A_{i,j}}z_2/z_1)
f_i(z_1,p)f_j(z_2,p)\nonumber\\
&&=-z_2 \Theta_{p}(q^{-A_{j,i}}z_2/z_1) 
f_j(z_2,p)f_i(z_1,p),~~{\rm for}~|A_{i,j}|\neq 0,\\
&&
~[f_i(z_1,p),f_j(z_2,p)]=0,~~{\rm for}~|A_{i,j}|=0, 
(i,j)\neq (M,M),\\
&&
\{f_M(z_1,p),f_M(z_2,p)\}=0,
\\
&&~[e_i(z_1,p),f_j(z_2,p)]
=\frac{\delta_{i,j}}{(q-q^{-1})z_1z_2}
\left(
\delta(q^{-c}z_1/z_2)\psi_i^+(q^{\frac{c}{2}}z_2,p)-
\delta(q^{c}z_1/z_2)\psi_i^-(q^{-\frac{c}{2}}z_2,p)
\right),\nonumber\\
&&
~~~~~{\rm for}~(i,j)\neq (M,M),
\\
&&~\{e_M(z_1,p),f_M(z_2,p)\}
=\frac{1}{(q-q^{-1})z_1z_2}
\left(
\delta(q^{-c}z_1/z_2)\psi_M^+(q^{\frac{c}{2}}z_2,p)-
\delta(q^{c}z_1/z_2)\psi_M^-(q^{-\frac{c}{2}}z_2,p)
\right),\nonumber\\
\end{eqnarray}
\begin{eqnarray}
&&~~\left(
e_i(z_1,p)e_i(z_2,p)e_j(z,p)
\frac{
\{q^{A_{i,j}}\frac{z}{z_1}\}^* 
\{q^{A_{i,j}}\frac{z}{z_2}\}^*}{
\{q^{-A_{i,j}}\frac{z}{z_1}\}^*
\{q^{-A_{i,j}}\frac{z}{z_2}\}^*}\right.\nonumber\\
&&-(q+q^{-1})
e_i(z_1,p)e_j(z,p)e_i(z_2,p)
\frac{
\{q^{A_{i,j}}\frac{z}{z_1}\}^* 
\{q^{A_{i,j}}\frac{z_2}{z}\}^*}{
\{q^{-A_{i,j}}\frac{z}{z_1}\}^* 
\{q^{-A_{i,j}}\frac{z_2}{z}\}^*}
\nonumber
\\
&&\left.
+e_j(z,p)e_i(z_1,p)e_i(z_2,p)
\frac{
\{q^{A_{i,j}}z_1/z\}^* 
\{q^{A_{i,j}}z_2/z\}^*}{
\{q^{-A_{i,j}}\frac{z_1}{z}\}^* 
\{q^{-A_{i,j}}\frac{z_2}{z}\}^*}
\right)
\frac{\{q^{A_{i,i}}\frac{z_2}{z_1}\}^*}{
\{q^{-A_{i,i}}\frac{z_2}{z_1}\}^*}\nonumber\\
&&+(z_1 \leftrightarrow z_2)=0,~~
{\rm for}~|A_{i,j}|=1, i \neq M,
\\
&&~~\left(f_i(z_1,p)f_i(z_2,p)f_j(z,p)
\frac{
\{q^{-A_{i,j}}\frac{z}{z_1}\} 
\{q^{-A_{i,j}}\frac{z}{z_2}\}}{
\{q^{A_{i,j}}\frac{z}{z_1}\} 
\{q^{A_{i,j}}\frac{z}{z_2}\}}\right.\nonumber\\
&&
\left.
-(q+q^{-1})
f_i(z_1,p)f_j(z,p)f_i(z_2,p)
\frac{
\{q^{-A_{i,j}}\frac{z}{z_1}\} 
\{q^{-A_{i,j}}\frac{z_2}{z}\}}{
\{q^{A_{i,j}}\frac{z}{z_1}\} 
\{q^{A_{i,j}}\frac{z_2}{z}\}}\right.
\nonumber
\\
&&\left.
+
f_j(z,p)f_i(z_1,p)f_i(z_2,p)
\frac{
\{q^{-A_{i,j}}\frac{z_1}{z}\} 
\{q^{-A_{i,j}}\frac{z_2}{z}\}}{
\{q^{A_{i,j}}\frac{z_1}{z}\} 
\{q^{A_{i,j}}\frac{z_2}{z}\}}
\right)
\frac{\{q^{-A_{i,i}}\frac{z_2}{z_1}\}}{
\{q^{A_{i,i}}\frac{z_2}{z_1}\}}\nonumber\\
&&
+(z_1 \leftrightarrow z_2)=0,
~~{\rm for}~|A_{i,j}|=1, i\neq M,
\end{eqnarray}
\begin{eqnarray}
&&
\left(
e_M(z_1,p)e_{M+1}(w_1,p)e_M(z_2,p)e_{M-1}(w_2,p)
\frac{
\{\frac{qw_1}{z_1}\}^*
\{\frac{qz_2}{w_1}\}^*
\{\frac{w_2}{qz_1}\}^*
\{\frac{w_2}{qz_2}\}^*}
{
\{\frac{w_1}{qz_1}\}^*
\{\frac{z_2}{qw_1}\}^*
\{\frac{qw_2}{z_1}\}^*
\{\frac{qw_2}{z_2}\}^*
}\right.\nonumber\\
&&
-q^{-1}
e_M(z_1,p)e_{M+1}(w_1,p)e_{M-1}(w_2,p)e_M(z_2,p)
\frac{
\{\frac{qw_1}{z_1}\}^*
\{\frac{w_2}{qz_1}\}^*
\{\frac{qz_2}{w_1}\}^*
\{\frac{z_2}{qw_2}\}^*}
{
\{\frac{w_1}{qz_1}\}^*
\{\frac{qw_2}{z_1}\}^*
\{\frac{z_2}{qw_1}\}^*
\{\frac{qz_2}{w_2}\}^*
}\nonumber\\
&&-q
e_M(z_1,p)
e_M(z_2,p)
e_{M-1}(w_2,p)
e_{M+1}(w_1,p)
\frac{
\{\frac{w_2}{qz_1}\}^*
\{\frac{w_2}{qz_2}\}^*
\{\frac{qw_1}{z_1}\}^*
\{\frac{qw_1}{z_2}\}^*}
{
\{\frac{qw_2}{z_2}\}^*
\{\frac{qw_1}{z_1}\}^*
\{\frac{w_1}{qz_2}\}^*
\{\frac{z_2}{qw_2}\}^*
}
\nonumber\\
&&
+
e_M(z_1,p)
e_{M-1}(w_2,p)
e_M(z_2,p)
e_{M+1}(w_1,p)
\frac{
\{\frac{w_2}{qz_1}\}^*
\{\frac{z_2}{qz_1}\}^*
\{\frac{qw_1}{z_1}\}^*
\{\frac{qw_1}{z_2}\}^*}
{
\{\frac{qw_2}{z_1}\}^*
\{\frac{qz_2}{z_1}\}^*
\{\frac{w_1}{qz_1}\}^*
\{\frac{w_1}{qz_2}\}^*
}
\nonumber\\
&&
+e_{M+1}(w_1,p)
e_{M}(z_2,p)
e_{M-1}(w_2,p)
e_{M}(z_1,p)
\frac{
\{\frac{qz_2}{w_1}\}^*
\{\frac{w_2}{qz_2}\}^*
\{\frac{qz_1}{w_1}\}^*
\{\frac{z_1}{qw_2}\}^*}
{
\{\frac{z_2}{qw_1}\}^*
\{\frac{qw_2}{z_2}\}^*
\{\frac{z_1}{qw_1}\}^*
\{\frac{qz_1}{w_2}\}^*
}
\nonumber\\
&&-q^{-1}
e_{M+1}(w_1,p)
e_{M-1}(w_2,p)
e_{M}(z_2,p)
e_{M}(z_1,p)
\frac{
\{\frac{qz_2}{w_1}\}^*
\{\frac{z_2}{qw_2}\}^*
\{\frac{qz_1}{w_1}\}^*
\{\frac{z_1}{qw_2}\}^*}
{
\{\frac{z_2}{qw_1}\}^*
\{\frac{qz_2}{w_2}\}^*
\{\frac{z_1}{qw_1}\}^*
\{\frac{qz_1}{w_2}\}^*
}
\nonumber\\
&&
-q
e_{M}(z_2,p)
e_{M-1}(w_2,p)
e_{M+1}(w_1,p)
e_{M}(z_1,p)
\frac{
\{\frac{w_2}{qz_2}\}^*
\{\frac{w_1}{qz_2}\}^*
\{\frac{qz_1}{w_2}\}^*
\{\frac{qz_1}{w_1}\}^*}
{
\{\frac{qw_2}{z_2}\}^*
\{\frac{qw_1}{z_2}\}^*
\{\frac{z_1}{qw_2}\}^*
\{\frac{z_1}{qw_1}\}^*}
\nonumber\\
&&
+\left.
e_{M-1}(w_2,p)
e_{M}(z_2,p)
e_{M+1}(w_1,p)
e_{M}(z_1,p)
\frac{
\{\frac{z_2}{qw_2}\}^*
\{\frac{qw_1}{z_1}\}^*
\{\frac{z_1}{qw_2}\}^*
\{\frac{qz_1}{w_1}\}^*}
{
\{\frac{qz_2}{w_2}\}^*
\{\frac{w_1}{qz_1}\}^*
\{\frac{qz_1}{w_2}\}^*
\{\frac{z_1}{qw_1}\}^*
}
\right)\nonumber\\
&&+(z_1 \leftrightarrow z_2)=0,
\end{eqnarray}
\begin{eqnarray}
&&
\left(
f_M(z_1,p)f_{M+1}(w_1,p)f_M(z_2,p)f_{M-1}(w_2,p)
\frac{
\{\frac{w_1}{qz_1}\}
\{\frac{z_2}{qw_1}\}
\{\frac{qw_2}{z_1}\}
\{\frac{qw_2}{z_2}\}}
{
\{\frac{qw_1}{z_1}\}
\{\frac{qz_2}{w_1}\}
\{\frac{w_2}{qz_1}\}
\{\frac{w_2}{qz_2}\}
}\right.\nonumber\\
&&
-q^{-1}
f_M(z_1,p)
f_{M+1}(w_1,p)
f_{M-1}(w_2,p)
f_M(z_2,p)
\frac{
\{\frac{w_1}{qz_1}\}
\{\frac{qw_2}{z_1}\}
\{\frac{z_2}{qw_1}\}
\{\frac{qz_2}{w_2}\}}
{
\{\frac{qw_1}{z_1}\}
\{\frac{w_2}{qz_1}\}
\{\frac{qz_2}{w_1}\}
\{\frac{z_2}{qw_2}\}
}\nonumber\\
&&-q
f_M(z_1,p)
f_M(z_2,p)
f_{M-1}(w_2,p)
f_{M+1}(w_1,p)
\frac{
\{\frac{qw_2}{z_1}\}
\{\frac{qw_2}{z_2}\}
\{\frac{w_1}{qz_1}\}
\{\frac{w_1}{qz_2}\}}
{
\{\frac{w_2}{qz_2}\}
\{\frac{w_1}{qz_1}\}
\{\frac{qw_1}{z_2}\}
\{\frac{qz_2}{w_2}\}
}
\nonumber\\
&&
+
f_M(z_1,p)
f_{M-1}(w_2,p)
f_M(z_2,p)
f_{M+1}(w_1,p)
\frac{
\{\frac{qw_2}{z_1}\}
\{\frac{qz_2}{z_1}\}
\{\frac{w_1}{qz_1}\}
\{\frac{w_1}{qz_2}\}}
{
\{\frac{w_2}{qz_1}\}
\{\frac{z_2}{qz_1}\}
\{\frac{qw_1}{z_1}\}
\{\frac{qw_1}{z_2}\}
}
\nonumber\\
&&
+
f_{M+1}(w_1,p)
f_{M}(z_2,p)
f_{M-1}(w_2,p)
f_{M}(z_1,p)
\frac{
\{\frac{z_2}{qw_1}\}
\{\frac{qw_2}{z_2}\}
\{\frac{z_1}{qw_1}\}
\{\frac{qz_1}{w_2}\}}
{
\{\frac{qz_2}{w_1}\}
\{\frac{w_2}{qz_2}\}
\{\frac{qz_1}{w_1}\}
\{\frac{z_1}{qw_2}\}
}
\nonumber\\
&&-q^{-1}
f_{M+1}(w_1,p)
f_{M-1}(w_2,p)
f_{M}(z_2,p)
f_{M}(z_1,p)
\frac{
\{\frac{z_2}{qw_1}\}
\{\frac{qz_2}{w_2}\}
\{\frac{z_1}{qw_1}\}
\{\frac{qz_1}{w_2}\}}
{
\{\frac{qz_2}{w_1}\}
\{\frac{z_2}{qw_2}\}
\{\frac{qz_1}{w_1}\}
\{\frac{z_1}{qw_2}\}
}
\nonumber\\
&&
-q
f_{M}(z_2,p)
f_{M-1}(w_2,p)
f_{M+1}(w_1,p)
f_{M}(z_1,p)
\frac{
\{\frac{qw_2}{z_2}\}
\{\frac{qw_1}{z_2}\}
\{\frac{z_1}{qw_2}\}
\{\frac{z_1}{qw_1}\}}
{
\{\frac{w_2}{qz_2}\}
\{\frac{w_1}{qz_2}\}
\{\frac{qz_1}{w_2}\}
\{\frac{qz_1}{w_1}\}}
\nonumber\\
&&
+\left.
f_{M-1}(w_2,p)
f_{M}(z_2,p)
f_{M+1}(w_1,p)
f_{M}(z_1,p)
\frac{
\{\frac{qz_2}{w_2}\}
\{\frac{w_1}{qz_1}\}
\{\frac{qz_1}{w_2}\}
\{\frac{z_1}{qw_1}\}}
{
\{\frac{z_2}{qw_2}\}
\{\frac{qw_1}{z_1}\}
\{\frac{z_1}{qw_2}\}
\{\frac{qz_1}{w_1}\}
}
\right)\nonumber\\
&&+(z_1 \leftrightarrow z_2)=0.
\end{eqnarray}
We have used the abbreviations
(\ref{def:abb}).
\end{prop}

\begin{prop}~~~
The currents
$\psi_j^\pm(z)$ $(1 \leqq j \leqq M+N-1)$ have
the following formulae.
\begin{eqnarray}
\psi_j^\pm(q^{\mp (r-\frac{c}{2})}z,p)=q^{\mp h_j}
:\exp\left(-\sum_{m \neq 0}\frac{B_{j,m}}{[r^*m]_q}
z^{-m}\right):.
\end{eqnarray}
Here we have set
\begin{eqnarray}
B_{j,m}=\left\{
\begin{array}{cc}
\frac{[r^*m]_q}{[rm]_q}a_{j,m},&~~(m>0)\\
q^{c|m|}a_{j,m},&~~(m<0)
\end{array}
\right.~~~(1\leqq j \leqq M+N-1).
\end{eqnarray}
\end{prop}

\begin{dfn}~~~
We define elliptic currents
$E_j(z), F_j(z), H_j(z), (1\leqq j \leqq M+N-1)$ by
\begin{eqnarray}
&&E_j(z)=e_j(z,p)e^{2Q_j}z^{-\frac{1}{r^*}P_j},\\
&&F_j(z)=f_j(z,p)z^{\frac{1}{r}(P_j+h_j)},\\
&&H_j^\pm(z)=H_j(q^{\pm(r-\frac{c}{2})}z),\\
&&H_j(z)=:\exp\left(-\sum_{m\neq 0}
\frac{B_{j,m}}{[r^*m]_q}z^{-m}
\right):e^{2Q_j}z^{-\frac{c}{r r^*}P_j+\frac{1}{r}h_j}.
\end{eqnarray}
Here we have used the zero-mode operators $P_j, Q_j$,
$(1\leqq j \leqq M+N-1)$.
\begin{eqnarray}
[P_i,Q_j]=-\frac{A_{i,j}}{2},~~(1\leqq i,j \leqq M+N-1).
\end{eqnarray}
\end{dfn}

\begin{prop}~~~
The currents $E_j(z), F_j(z), H_j(z)$ and $B_{j,n},
h_j, c$, $(1\leqq j \leqq M+N-1, n \in {\mathbb Z}_{\neq 0})$ satisfy
the defining relations
of elliptic superalgebra 
$U_{q,p}(\widehat{sl}(M|N))$
(\ref{def:elliptic1}), 
(\ref{def:elliptic2}), 
(\ref{def:elliptic3}), 
(\ref{def:elliptic4}), 
(\ref{def:elliptic5}), 
(\ref{def:elliptic6}), 
(\ref{def:elliptic7}), 
(\ref{def:elliptic8}), 
(\ref{def:elliptic9}), 
(\ref{def:elliptic10}), 
(\ref{def:elliptic11}), 
(\ref{def:elliptic12}).
They satisfy
the Serre relations
(\ref{eqn:Serre1}), (\ref{eqn:Serre2})
and
(\ref{eqn:Serre3}),(\ref{eqn:Serre4})
for $1\leqq i,j \leqq M+N-1$,$(i\neq M)$ 
such that $|A_{i,j}|=1$.

\end{prop}
We have constructed the elliptic deformed superalgebra
$U_{q,p}(\widehat{sl}(M|N))$ from
the quantum superalgebra $U_q(\widehat{sl}(M|N))$.

\section{Bosonization}

In this section
we give new bosonization
of the superalgebra
$U_{q}(\widehat{sl}(1|2)), U_{q,p}(\widehat{sl}(1|2))$
for an arbitrary level $k$,
and their screening currents.
Wakimoto \cite{Wakimoto} constructed bosonization
of affine algebra $\widehat{sl}(2)$ for an arbitrary 
level $k$. We call this-type bosonization based on the flag manifold \cite{FF2}
the Wakimoto realization.
Feigin-Frenkel \cite{FF} generalized 
the Wakimoto realization
to the higher-rank affine algebra $\widehat{sl}(N)$.
Shiraishi \cite{Shiraishi}
constructed the Wakimoto realization
of the quantum algebra $U_q(\widehat{sl}(2))$
and its screening currents.
Awata-Odake-Shiraishi
constructed the Wakimoto realization for 
the quantum algebra $U_q(\widehat{sl}(N))$ and its
screening currents \cite{AOS1}.
In the case of $U_q(\widehat{sl}(2|1))$
Awata-Odake-Shiraishi \cite{AOS2}
constructed the Wakimoto realization and
Zhang-Gould \cite{ZG2}
constructed the screening currents.

\subsection{
$U_q(\widehat{sl}(1|2))$, 
$U_{q,p}(\widehat{sl}(1|2))$, Screening}

In this section we give new bosonizations of
$U_{q}(\widehat{sl}(1|2))$,
$U_{q,p}(\widehat{sl}(1|2))$
and their screening currents.
In this section we assume 
the central element $c=k\neq 1$.
The Cartan matrix $(A_{i,j})_{0\leqq i,j \leqq 2}$
of $\widehat{sl}(1|2)$ is given by
\begin{eqnarray}
(A_{i,j})_{0\leqq i,j \leqq 2}
=\left(\begin{array}{ccc}
0&-1&1\\
-1&0&1\\
1&1&-2
\end{array}\right).
\end{eqnarray}
The Cartan matrix of the classical part
$sl(1|2)$ is written by
\begin{eqnarray}
(A_{i,j})_{1\leqq i,j \leqq 2}=
((\nu_i+\nu_{i+1})\delta_{i,j}-\nu_i \delta_{i,j+1}
-\nu_{i+1}\delta_{i+1,j})_{1\leqq i,j \leqq 2},\nonumber
\end{eqnarray}
where we have set
$\nu_1=+, \nu_2=\nu_3=-$.
Let us introduce the bosons 
and the zero-mode operators
$a_m^j, Q_a^j$ $(m \in {\mathbb Z},
j=1,2)$ $b_m^{i,j}, Q_b^{i,j}$,
$c_m^{i,j}, Q_c^{i,j}$
$(m \in {\mathbb Z}, 1\leqq i<j \leqq 3)$ by
\begin{eqnarray}
&&~[a_m^i,a_n^j]=\frac{[(k-1)m]_q[A_{i,j}m]_q}{m}
\delta_{m+n,0},
~~[a_m^i, Q_a^j]=(k-1)A_{i,j}\delta_{m,0}, 
\label{def:boson1}
\\
&&~[b_m^{i,j},b_n^{i',j'}]=
-\nu_i \nu_j\frac{[m]_q^2}{m}
\delta_{i,i'}\delta_{j,j'}\delta_{m+n,0},
~~[b_m^{i,j},Q_b^{i',j'}]=
-\nu_i \nu_j\delta_{i,i'}
\delta_{j,j'}\delta_{m,0},
\\
&&~[c_m^{i,j},c_n^{i',j'}]=
\nu_i \nu_j\frac{[m]_q^2}{m}
\delta_{i,i'}\delta_{j,j'}
\delta_{m+n,0},
~~[c_m^{i,j},Q_c^{i',j'}]=
\nu_i \nu_j \delta_{i,i'}\delta_{j,j'}
\delta_{m,0}.
\end{eqnarray}
Let us set the bosonic fields $a(z)$, $a_\pm(z)$
and $\left(\frac{1}{\beta}~a\right)(z|\alpha)$
as follows.
\begin{eqnarray}
&&a(z)=-\sum_{m \neq 0}\frac{a_m}{[m]_q}z^{-m}+Q_a+
a_0{\rm log}z,\\
&&a_\pm(z)=\pm (q-q^{-1})\sum_{m>0}a_{\pm m}z^{\mp m}
\pm a_0 {\rm log}q,\\
&&\left(\frac{1}{\beta}~a\right)
(z|\alpha)=-\sum_{m \neq 0}
\frac{a_m}{[\beta m]}q^{-\alpha |m|}z^{-m}+
\frac{1}{\beta}(Q_a+a_0 {\rm log}z).
\end{eqnarray}
We impose the
cocycle condition to the zero-mode operator. 
\begin{eqnarray}
e^{Q_b^{1,2}}e^{Q_b^{1,3}}=-e^{Q_b^{1,3}}e^{Q_b^{1,2}},~
e^{Q_b^{1,2}}e^{Q_b^{2,3}}=e^{Q_b^{2,3}}e^{Q_b^{1,2}},~
e^{Q_b^{1,2}}e^{Q_b^{2,3}}=e^{Q_b^{2,3}}e^{Q_b^{1,3}}.
\end{eqnarray}
Straightforward OPE calculations show the following propositions.

\begin{prop}~~~
Bosonization of the quantum superalgebra
$U_{q,p}(\widehat{sl}(1|2))$
is given as follows.
\begin{eqnarray}
&&c=k,~~h_1=a_0^1-b_0^{2,3}-b_0^{1,2},~~
h_2=a_0^2+2b_0^{2,3}+b_0^{1,3}-b_0^{1,2},\\
&&a_{1,m}=a_m^1q^{-\frac{k-1}{2}|m|}
-b_m^{2,3}q^{-(k-1)|m|}
-b_m^{1,3}q^{-(k-1)|m|},\\
&&a_{2,m}=a_m^2 q^{-\frac{k-1}{2}|m|}
+b_m^{2,3}q^{-(k-1)|m|}(q^m+q^{-m})
+b_m^{1,3}q^{-(k-2)|m|}
-b_m^{1,2}q^{-(k-1)|m|},\\
&&x^+_1(z)=c_{1,1}^+ x_{1,1}^+(z)+c_{1,2}^+ x_{1,2}^+(z),
\label{boson:q1}\\
&&x^+_2(z)=\frac{1}{(q-q^{-1})z}(
c_{2,1}^+ x_{2,1}^+(z)-c_{2,2}^+ x_{2,2}^+(z)),
\label{boson:q2}\\
&&x^-_1(z)=\frac{1}{(q-q^{-1})z}
(c_{1,1}^- x_{1,1}^-(z)-c_{1,2}^- x_{1,2}^-(z)
-c_{1,3}^- x_{1,3}^-(z)+c_{1,4}^- x_{1,4}^-(z)),
\label{boson:q3}\\
&&x^-_2(z)=\frac{1}{(q-q^{-1})z}(
c_{2,1}^- x_{2,1}^-(z)-
c_{2,2}^- x_{2,2}^-(z))+c_{2,3}^- x_{2,3}^-(z),
\label{boson:q4}
\end{eqnarray}
where we have set
\begin{eqnarray}
&&x_{1,1}^+(z)=:e^{-(b^{2,3}+b^{1,3})_+(q^{-1}z)
-b^{1,2}(q^{-1}z)}:,\\
&&x_{1,2}^+(z)=:e^{-(b+c)^{2,3}(z)-b^{1,3}(z)}:,\\
&&x_{2,1}^+(z)=:e^{b_+^{2,3}(z)+(b+c)^{2,3}(q^{-1}z)}:,\\
&&x_{2,2}^+(z)=:e^{b_-^{2,3}(z)+(b+c)^{2,3}(qz)}:,\\
&&x_{1,1}^-(z)=:e^{a_+^1(q^{\frac{k-1}{2}}z)
+b^{1,2}(q^{k-1}z)}:,\\
&&x_{1,2}^-(z)=:e^{a_-^1(q^{-\frac{k-1}{2}}z)
+b^{1,2}(q^{-k+1}z)}:,\\
&&x_{1,3}^-(z)=:e^{a_-^1(q^{-\frac{k-1}{2}}z)-b_-^{2,3}(q^{-k+1}z)
+(b+c)^{2,3}(q^{-k}z)
-b_-^{1,3}(q^{-k+1}z)+b^{1,3}(q^{-k}z)}:,\\
&&x_{1,4}^-(z)=:e^{a_-^1(q^{-\frac{k-1}{2}}z)
-b^{2,3}_+(q^{-k+1}z)-b_-^{1,3}(q^{-k+1}z)
+(b+c)^{2,3}(q^{-k+2}z)+b^{1,3}(q^{-k}z)}:,\\
&&x_{2,1}^-(z)=:e^{a_+^2(q^{\frac{k-1}{2}}z)+b_+^{2,3}(q^{k-2}z)
-(b+c)^{2,3}(q^{k-1}z)+b_+^{1,3}(q^{k-2}z)-b_+^{1,2}(q^{k-1}z)}:,\\
&&x_{2,2}^-(z)=:e^{a_-^2(q^{-\frac{k-1}{2}}z)
+b_-^{2,3}(q^{-k+2}z)
-(b+c)^{2,3}(q^{-k+1}z)
+b_-^{1,3}(q^{-k+2}z)-b_-^{1,2}(q^{-k+1}z)}:,\\
&&x_{2,3}^-(z)=
:e^{a_+^2(q^{\frac{k-1}{2}}z)+b^{1,3}(q^{k-1}z)
-b_+^{1,2}(q^{k-1}z)-b^{1,2}(q^{k-2}z)}:.
\end{eqnarray}
Here we have set the coefficients
as follows.
\begin{eqnarray}
&&(c_{1,1}^+,c_{1,2}^+,c_{2,1}^+,c_{2,2}^+)=
\left(\alpha,\beta,\gamma,\gamma\right),
\label{def:coef1(1|2)}
\\
&&(c_{1,1}^-,
c_{1,2}^-,c_{1,3}^-,c_{1,4}^-,
c_{2,1}^-,c_{2,2}^-,c_{2,3}^-)=
\left(\frac{1}{q\alpha},
\frac{1}{q \alpha},
\frac{1}{\beta},
\frac{1}{\beta},
\frac{1}{\gamma},
\frac{1}{\gamma},
\frac{q^{k-1}\alpha}{
\beta \gamma}\right).
\label{def:coef2(1|2)}
\end{eqnarray}
Here $\alpha,\beta,\gamma \neq 0$ are arbitrary
parameters.
\end{prop}

~\\
Next we give bosonization
of the elliptic superalgebra 
$U_{q,p}(\widehat{sl}(1|2))$.
Our construction is based on the dressing procedure
of the quantum algebra
developed in this paper.

\begin{prop}~~Bosonization of the elliptic
superalgebra $U_{q,p}(\widehat{sl}(1|2))$
is given as follows.
\begin{eqnarray}
&&c=k,~~h_1=a_0^1-b_0^{2,3}-b_0^{1,2},~~
h_2=a_0^2+2b_0^{2,3}+b_0^{1,3}-b_0^{1,2},\\
&&B_{j,m}=
\left\{
\begin{array}{cc}
\frac{[r^*m]_q}{[rm]_q}a_{j,m},&~~(m>0)\\
q^{k|m|}a_{j,m},&~~(m<0)
\end{array}
\right.,~~(j=1,2),
\\
&&a_{1,m}=a_m^1q^{-\frac{k-1}{2}|m|}
-b_m^{2,3}q^{-(k-1)|m|}
-b_m^{1,3}q^{-(k-1)|m|},
\\
&&a_{2,m}=a_m^2 q^{-\frac{k-1}{2}|m|}
+b_m^{2,3}q^{-(k-1)|m|}(q^m+q^{-m})
+b_m^{1,3}q^{-(k-2)|m|}
-b_m^{1,2}q^{-(k-1)|m|},
\\
&&E_j(z)=u_j^+(z,p)x_j^+(z)e^{2Q_j}z^{-\frac{1}{r^*}P_j},
~(j=1,2),\\
&&F_j(z)=x_j^-(z)u_j^-(z,p)z^{\frac{1}{r}(P_j+h_j)},
~(j=1,2),\\
&&H_j^\pm(z)=H_j(q^{\pm(r-\frac{c}{2})}z),~(j=1,2),
\end{eqnarray}
where we have used (\ref{boson:q1}), (\ref{boson:q2}),
(\ref{boson:q3}), (\ref{boson:q4}) and
\begin{eqnarray}
&&u_j^+(z,p)=\exp\left(
\sum_{m>0}\frac{q^{r^* m}}{[r^* m]_q}B_{j,-m}z^m
\right),~(j=1,2),\\
&&u_j^-(z,p)=\exp\left(
-\sum_{m>0}\frac{q^{rm}}{[r^*m]_q}B_{j,m}z^{-m}
\right),~(j=1,2),\\
&&H_j(z)=:\exp\left(-\sum_{m\neq 0}
\frac{B_{j,m}}{[r^*m]_q}z^{-m}
\right):e^{2Q_j}z^{-\frac{c}{r r^*}P_j+\frac{1}{r}h_j},~(j=1,2).
\end{eqnarray}
Here we have used the zero-mode operators
\begin{eqnarray}
[P_i,Q_j]=-\frac{A_{i,j}}{2},~ (1\leqq i,j \leqq 2).
\end{eqnarray}
\end{prop}

\begin{prop}~~
The bosonic operators $s_j(z)~(j=1,2)$ given below
are the screening currents
that commute with the quantum superalgebra
$U_q(\widehat{sl}(1|2))$ modulo total difference.
\begin{eqnarray}
s_j(z)=:e^{-(\frac{1}{k-1}a^j)(z_1|\frac{k-1}{2})}
\tilde{s}_{j}(z):~~(j=1,2).\label{def:screening(1|2)}
\end{eqnarray}
Here we have set
\begin{eqnarray}
\tilde{s}_1(z)&=&-c_{1,5}\tilde{s}_{1,5}(z),
\label{def:screening1(1|2)}\\
\tilde{s}_2(z)&=&\frac{1}{(q-q^{-1})z}(
-c_{2,3}\tilde{s}_{2,3}(z)+c_{2,4}\tilde{s}_{2,4}(z))
+c_{2,5}\tilde{s}_{2,5}(z),
\label{def:screening2(1|2)}
\end{eqnarray}
where
\begin{eqnarray}
\tilde{s}_{1,5}(z)&=&:e^{-b^{1,2}(z)}:,\\
\tilde{s}_{2,3}(z)&=&
:e^{-b_+^{2,3}(qz)+(b+c)^{2,3}(q^2z)
-b_-^{1,3}(qz)+b_-^{1,2}(z)}:,\\
\tilde{s}_{2,4}(z)&=&:e^{-b_-^{2,3}(qz)+(b+c)^{2,3}(z)
-b_-^{1,3}(qz)+b_-^{1,2}(z)}:,\\
\tilde{s}_{2,5}(z)&=&:e^{-b^{1,3}(z)+b_+^{1,2}(z)
+b^{1,2}(q^{-1}z)}:.
\end{eqnarray}
Here we have set the coefficients as follows.
\begin{eqnarray}
&&(c_{1,5},c_{2,3},c_{2,4},c_{2,5})=\left(q\alpha,\gamma,\gamma,
\frac{\beta \gamma}{q \alpha}\right),
\end{eqnarray}
where parameters $\alpha,\beta,\gamma \neq 0$
have been introduced in
(\ref{def:coef1(1|2)}), (\ref{def:coef2(1|2)})
for the bosonizations of
$U_q(\widehat{sl}(1|2))$.
Explicitly the bosonic operators
$s_1(z), s_2(z)$ and $x_1^\pm(z),
x_2^\pm(z)$ satisfy the following relations.
\begin{eqnarray}
~[a_{i,m},s_j(z_2)]&=&0,\\
~[x_i^+(z_1),s_j(z_2)]&=&0,\\
~[x_i^-(z_1),s_j(z_2)]&=&
\frac{\delta_{i,j}}{(q-q^{-1})z_1z_2}
(\delta(q^{k-1}z_2/z_1)-
\delta(q^{-k+1}z_2/z_1))\nonumber\\
&\times&
:e^{-(\frac{1}{k-1}a^i)(z_1|-\frac{k-1}{2})}:,
\end{eqnarray}
and
\begin{eqnarray}
\{\tilde{s}_1(z_1), \tilde{s}_1(z_2)\}&=&0,\\
(z_1-q^{-A_{1,2}}z_2)\tilde{s}_1(z_1)\tilde{s}_2(z_2)
&=&
(q^{-A_{2,1}}z_1-z_2)\tilde{s}_2(z_2)\tilde{s}_1(z_1),\\
(z_1-q^{-A_{2,2}}z_2)\tilde{s}_2(z_1)\tilde{s}_2(z_2)
&=&
(q^{-A_{2,2}}z_1-z_2)\tilde{s}_2(z_2)\tilde{s}_2(z_1).
\end{eqnarray}
\end{prop}
By the commutation relation
$[a_{i,m},s_j(z_2)]=0$ we conclude the following.
\begin{prop}~~
The bosonic operators $s_j(z)~(j=1,2)$
given in (\ref{def:screening(1|2)})
become the screening currents
that commute with 
the elliptic algebra $U_{q,p}(\widehat{sl}(1|2))$
modulo total difference.
Explicitly the bosonic operators
$s_1(z), s_2(z)$ and $E_1(z),E_2(z),F_1(z),F_2(z)$ 
satisfy the following relations.
\begin{eqnarray}
~[B_{i,m},s_j(z_2)]&=&0,\\
~[E_i(z_1),s_j(z_2)]&=&0,\\
~[F_i(z_1),s_j(z_2)]&=&
\frac{\delta_{i,j}}{(q-q^{-1})z_1z_2}
(\delta(q^{k-1}z_2/z_1)-
\delta(q^{-k+1}z_2/z_1))\nonumber\\
&\times&
:e^{-(\frac{1}{k-1}a^i)(z_1|-\frac{k-1}{2})}
u_i^-(z_1,p)z_1^{\frac{1}{r}(P_i+h_i)}:.
\end{eqnarray}
\end{prop}
The Jackson integral with parameters $p$ and $s \neq 0$ is
defined by
\begin{eqnarray}
\int_0^{s \infty}f(z)d_pz=s(1-p)\sum_{n \in {\mathbb Z}}f(sp^n)p^n.
\end{eqnarray}
From the above proposition we have 
\begin{eqnarray}
\left[\int_0^{s \infty}s_j(z)d_{q^{2(k-1)}}z,~U_{q,p}(\widehat{sl}(1|2))
\right]=0.
\end{eqnarray}

\subsection{$U_q(\widehat{sl}(2|1))$, 
$U_{q,p}(\widehat{sl}(2|1))$, Screening}

In this section we review known results on
bosonization of $U_q(\widehat{sl}(2|1))$ \cite{AOS2}
and 
its screening currents \cite{ZG2}.
We give bosonizations of $U_{q,p}(\widehat{sl}(2|1))$
and its screenings.
In this section we assume
the central element $c=k\neq -1$.
The Cartan matrix $(A_{i,j})_{0\leqq i,j \leqq 2}$
of $\widehat{sl}(2|1)$ is given by
\begin{eqnarray}
(A_{i,j})_{0\leqq i,j \leqq 2}
=\left(\begin{array}{ccc}
0&-1&1\\
-1&2&-1\\
1&-1&0
\end{array}\right).
\end{eqnarray}
The Cartan matrix of the classical part $sl(2|1)$ is written by
\begin{eqnarray}
(A_{i,j})_{1\leqq i,j \leqq 2}=
((\nu_i+\nu_{i+1})\delta_{i,j}-\nu_i \delta_{i,j+1}
-\nu_{i+1}\delta_{i+1,j})_{1\leqq i,j \leqq 2},\nonumber
\end{eqnarray}
where we have set
$\nu_1=\nu_2=+, \nu_3=-$.
Let us introduce the bosons 
and the zero-mode operators
$a_m^j, Q_a^j$, $(m \in {\mathbb Z},
j=1,2)$ $b_m^{i,j}, Q_b^{i,j}$,
$c_m^{i,j}, Q_c^{i,j}$
$(m \in {\mathbb Z}, 1\leqq i<j \leqq 3)$ by
\begin{eqnarray}
&&~[a_m^i,a_n^j]=\frac{[(k+1)m]_q[A_{i,j}m]_q}{m}
\delta_{m+n,0},
~~[a_m^i, Q_a^j]=(k+1)A_{i,j}\delta_{m,0}, 
\label{def:boson1'}
\\
&&~[b_m^{i,j},b_n^{i',j'}]=
-\nu_i \nu_j\frac{[m]_q^2}{m}
\delta{i,i'}\delta_{j,j'}\delta_{m+n,0},
~~[b_m^{i,j},Q_b^{i',j'}]=
-\nu_i \nu_j\delta_{i,i'}\delta_{j,j'}\delta_{m,0},
\\
&&~[c_m^{i,j},c_n^{i',j'}]=
\nu_i \nu_j\frac{[m]_q^2}{m}\delta_{i,i'}\delta_{j,j'}
\delta_{m+n,0},
~~[c_m^{i,j},Q_c^{i',j'}]=\nu_i \nu_j 
\delta_{i,i'}\delta_{j,j'}
\delta_{m,0}.
\end{eqnarray}
We impose the
cocycle condition to the zero-mode operators.
\begin{eqnarray}
e^{Q_b^{1,2}}e^{Q_b^{1,3}}=e^{Q_b^{1,3}}e^{Q_b^{1,2}},~
e^{Q_b^{1,2}}e^{Q_b^{2,3}}=e^{Q_b^{2,3}}e^{Q_b^{1,2}},~
e^{Q_b^{1,3}}e^{Q_b^{2,3}}=-e^{Q_b^{2,3}}e^{Q_b^{1,3}}.
\end{eqnarray}

\begin{prop}~~\cite{AOS2}~~
Bosonization of the quantum superalgebra
$U_q(\widehat{sl}(2|1))$
is given as follows.
\begin{eqnarray}
&&c=k,
~~
h_1=a_0^1+2b_0^{1,2}+b_0^{1,3}-b_0^{2,3},~~
h_2=a_0^2-b_0^{1,2}-b_0^{1,3},\\
&&a_{1,m}=a_m^1 q^{-\frac{k+1}{2}|m|}
+b_m^{1,2}q^{-(k+1)|m|}(q^m+q^{-m})
+b_m^{1,3}q^{-(k+2)|m|}
-b_m^{2,3}q^{-(k+1)|m|},\\
&&a_{2,m}=a_m^2q^{-\frac{k+1}{2}|m|}
-b_m^{1,2}q^{-(k+1)|m|}
-b_m^{1,3}q^{-(k+1)|m|},
\end{eqnarray}
\begin{eqnarray}
&&
x^+_1(z)=
\frac{1}{(q-q^{-1})z}
(c_{1,1}^+x_{1,1}^+(z)-c_{1,2}^+x_{1,2}^+(z)),
\label{boson:q1'}\\
&&
x^+_2(z)=c_{2,1}^+x_{2,1}^+(z)+
c_{2,2}^+ x_{2,2}^+(z),
\label{boson:q2'}
\\
&&
x^-_1(z)=
\frac{1}{(q-q^{-1})z}(c_{1,1}^-x_{1,1}^-(z)
-c_{1,2}^-x_{1,2}^-(z))
+c_{1,3}^-x_{1,3}^-(z),
\label{boson:q3'}
\\
&&x^-_2(z)=
\frac{1}{(q-q^{-1})z}(
c_{2,1}^-x_{2,1}^-(z)
-c_{2,2}^-x_{2,2}^-(z)
-c_{2,3}^-x_{2,3}^-(z)
+c_{2,4}^-x_{2,4}^-(z)),
\label{boson:q4'}
\end{eqnarray}
where we have set
\begin{eqnarray}
&&x_{1,1}^+(z)=:e^{b_+^{1,2}(z)-(b+c)^{1,2}(qz)}:,\\
&&x_{1,2}^+(z)=:e^{b_-^{1,2}(z)-(b+c)^{1,2}(q^{-1}z)}:,\\
&&x_{2,1}^+(z)=:e^{-b_+^{1,2}(qz)-b_+^{1,3}(qz)+b^{2,3}(qz)}:,\\
&&x_{2,2}^+(z)=:e^{(b+c)^{1,2}(z)+b^{1,3}(z)}:,\\
&&x_{1,1}^-(z)=:
e^{a_+^1(q^{\frac{k+1}{2}}z)
+b_+^{1,2}(q^{k+2}z)+(b+c)^{1,2}(q^{k+1}z)
+b_+^{1,3}(q^{k+2}z)
-b_+^{2,3}(q^{k+1}z)}:,\\
&&x_{1,2}^-(z)=:e^{a_-^1(q^{-\frac{k+1}{2}}z)
+b_-^{1,2}(q^{-k-2}z)+(b+c)^{1,2}(q^{-k-1}z)
+b_-^{1,3}(q^{-k-2}z)
-b_-^{2,3}(q^{-k-1}z)}:,
\\
&&x_{1,3}^-(z)=:
e^{a_+^1(q^{\frac{k+1}{2}}z)-b_+^{2,3}(q^{k+1}z)
-b^{1,3}(q^{k+1}z)+b^{2,3}(q^{k+1}z)}:,\\
&&x_{2,1}^-(z)=:e^{a_+^2(q^{\frac{k+1}{2}}z)-b^{2,3}(q^{k+1}z)}:,\\
&&
x_{2,2}^-(z)=
:e^{a_-^2(q^{-\frac{k+1}{2}}z)-b^{2,3}(q^{-k-1}z)}:,\\
&&x_{2,3}^-(z)=:e^{a_-^2(q^{-\frac{k+1}{2}}z)
-b_-^{1,2}(q^{-k-1}z)-b_-^{1,3}(q^{-k-1}z)
-(b+c)^{1,2}(q^{-k}z)-b^{1,3}(q^{-k}z)}:,\\
&&x_{2,4}^-(z)=:
e^{a_-^2(q^{-\frac{k+1}{2}})
-b_+^{1,2}(q^{-k-1}z)-b_-^{1,3}(q^{-k-1}z)
-(b+c)^{1,2}(q^{-k-2}z)-b^{1,3}(q^{-k}z)}:.
\end{eqnarray}
Here we have set the coefficients
as follows.
\begin{eqnarray}
&&(c_{1,1}^+,c_{1,2}^+,c_{2,1}^+,c_{2,2}^+)=
\left(\alpha,\alpha,\beta,\gamma\right),
\label{def:coef1(2|1)}\\
&&(c_{1,1}^-,
c_{1,2}^-,c_{1,3}^-,
c_{2,1}^-,c_{2,2}^-,
c_{2,3}^-,c_{2,4}^-)=
\left(\frac{1}{\alpha},\frac{1}{\alpha},\frac{q^{k+1}\beta}{\alpha \gamma},
\frac{q}{\beta},\frac{q}{\beta},\frac{1}{\gamma},\frac{1}{\gamma}
\right).\label{def:coef2(2|1)}
\end{eqnarray}
Here $\alpha,\beta,\gamma \neq 0$ are arbitrary
parameters.
\end{prop}

~\\
{\it Note.~
The coefficients of
the currents $x_j^\pm(z)$ have 4 free parameters 
in \cite{AOS2}.
In this paper we have only three free parameters 
$\alpha, \beta, \gamma$, because we assume
the commutation relations 
(\ref{eqn:screening1(2|1)}), (\ref{eqn:screening2(2|1)}), 
(\ref{eqn:screening3(2|1)})
with the screening currents.}

~\\

\begin{prop}~~Bosonization of the elliptic
superalgebra $U_{q,p}(\widehat{sl}(2|1))$
is given as follows.
\begin{eqnarray}
&&c=k,
~~
h_1=a_0^1+2b_0^{1,2}+b_0^{1,3}-b_0^{2,3},~~
h_2=a_0^2-b_0^{1,2}-b_0^{1,3},\\
&&B_{j,m}=
\left\{
\begin{array}{cc}
\frac{[r^*m]_q}{[rm]_q}a_{j,m},&~~(m>0)\\
q^{k|m|}a_{j,m},&~~(m<0)
\end{array}
\right.,~~(j=1,2),\\
&&a_{1,m}=a_m^1 q^{-\frac{k+1}{2}|m|}
+b_m^{1,2}q^{-(k+1)|m|}(q^m+q^{-m})
+b_m^{1,3}q^{-(k+2)|m|}
-b_m^{2,3}q^{-(k+1)|m|},\\
&&a_{2,m}=a_m^2q^{-\frac{k+1}{2}|m|}
-b_m^{1,2}q^{-(k+1)|m|}
-b_m^{1,3}q^{-(k+1)|m|},
\\
&&E_j(z)=u_j^+(z,p)x_j^+(z)e^{2Q_j}z^{-\frac{1}{r^*}P_j},
~(j=1,2),\\
&&F_j(z)=x_j^-(z)u_j^-(z,p)z^{\frac{1}{r}(P_j+h_j)},~
(j=1,2),\\
&&H_j^\pm(z)=H_j(q^{\pm(r-\frac{c}{2})}z),~(j=1,2),
\end{eqnarray}
where we have used (\ref{boson:q1'}), 
(\ref{boson:q2'}),
(\ref{boson:q3'}), 
(\ref{boson:q4'}) and
\begin{eqnarray}
&&u_j^+(z,p)=\exp\left(
\sum_{m>0}\frac{q^{r^* m}}{[r^* m]_q}B_{j,-m}z^m
\right),~(j=1,2),\\
&&u_j^-(z,p)=\exp\left(
-\sum_{m>0}\frac{q^{rm}}{[r^*m]_q}B_{j,m}z^{-m}
\right),~(j=1,2),\\
&&H_j(z)=:\exp\left(-\sum_{m\neq 0}
\frac{B_{j,m}}{[r^*m]_q}z^{-m}
\right):e^{2Q_j}z^{-\frac{c}{r r^*}P_j+\frac{1}{r}h_j},
~(j=1,2).
\end{eqnarray}
Here we have used the zero-mode operators
\begin{eqnarray}
[P_i,Q_j]=-\frac{A_{i,j}}{2},~ (1\leqq i,j \leqq 2).
\end{eqnarray}
\end{prop}

\begin{prop}~~\cite{ZG2}~~
The bosonic operators $s_1(z), s_2(z)$ given below
are the screening currents
that commute with the quantum superalgebra
$U_q(\widehat{sl}(2|1))$ modulo total difference.
\begin{eqnarray}
s_{j}(z)&=&
:e^{-(\frac{1}{k+1}a^j)
(z|\frac{k+1}{2})} \tilde{s}_{j}(z):~~(j=1,2).
\label{def:screening(2|1)}
\end{eqnarray}
Here we have set
\begin{eqnarray}
\tilde{s}_1(z)&=&\frac{1}{(q-q^{-1})z}
(-c_{1,3} \tilde{s}_{1,3}(z)+c_{1,4} \tilde{s}_{1,4}(z))
+c_{1,5} \tilde{s}_{1,5}(z),
\label{def:screening1(2|1)}
\\
\tilde{s}_2(z)&=&-c_{2,5} \tilde{s}_{2,5}(z),
\label{def:screening2(2|1)}
\end{eqnarray}
where
\begin{eqnarray}
\tilde{s}_{1,5}(z)&=&:e^{b^{1,3}(z)-b^{2,3}(qz)+b_+^{2,3}(z)}:,
\\
\tilde{s}_{1,4}(z)&=&:e^{-b_-^{1,2}(q^{-1}z)-(b+c)^{1,2}(z)
+b_-^{2,3}(z)-b_-^{1,3}(q^{-1}z)}:,
\\
\tilde{s}_{1,3}(z)&=&:e^{-b_+^{1,2}(q^{-1}z)-(b+c)^{1,2}(q^{-2}z)
+b_-^{2,3}(z)-b_-^{1,3}(q^{-1}z)}:,
\\
\tilde{s}_{2,5}(z)&=&:e^{b^{2,3}(z)}:.\nonumber
\end{eqnarray}
Here we have set the coefficients as follows.
\begin{eqnarray}
&&(c_{1,3},c_{1,4},c_{1,5},c_{2,5})=\left(\alpha,\alpha,
\frac{q \alpha \beta}{\gamma},\frac{\beta}{q}\right),
\end{eqnarray}
where parameters $\alpha,\beta,\gamma \neq 0$
have been introduced in
(\ref{def:coef1(2|1)}), (\ref{def:coef2(2|1)}) 
for the bosonizations of
$U_q(\widehat{sl}(2|1))$.
Explicitly the bosonic operators
$s_1(z), s_2(z)$ and $x_1^\pm(z),
x_2^\pm(z)$ satisfy the following relations.
\begin{eqnarray}
~[a_{i,m},s_j(z_2)]&=&0,
\label{eqn:screening1(2|1)}\\
~[x_i^+(z_1),s_j(z_2)]&=&0,
\label{eqn:screening2(2|1)}\\
~[x_i^-(z_1),s_j(z_2)]&=&
\frac{\delta_{i,j}}{(q-q^{-1})z_1z_2}
(\delta(q^{k+1}z_2/z_1)-
\delta(q^{-k-1}z_2/z_1))\nonumber\\
&\times&
:e^{-(\frac{1}{k+1}a^i)(z_1|-\frac{k+1}{2})}:,
\label{eqn:screening3(2|1)}
\end{eqnarray}
and
\begin{eqnarray}
(z_1-q^{-A_{1,1}}z_2)\tilde{s}_1(z_1)\tilde{s}_1(z_2)
&=&
(q^{-A_{1,1}}z_1-z_2)\tilde{s}_1(z_2)\tilde{s}_1(z_1),\\
(z_1-q^{-A_{1,2}}z_2)\tilde{s}_1(z_1)\tilde{s}_2(z_2)
&=&
(q^{-A_{2,1}}z_1-z_2)\tilde{s}_2(z_2)\tilde{s}_1(z_1),\\
\{\tilde{s}_2(z_1), \tilde{s}_2(z_2)\}&=&0.
\end{eqnarray}
\end{prop}
By the commutation relation
$[a_{i,m},s_j(z_2)]=0$ we conclude the following.
\begin{prop}~~
The bosonic operators $s_j(z)~(j=1,2)$ given 
in (\ref{def:screening(2|1)})
become the screening currents
that commute with 
the elliptic algebra $U_{q,p}(\widehat{sl}(2|1))$
modulo total difference.
Explicitly the bosonic operators
$s_1(z), s_2(z)$ and $E_1(z),E_2(z),F_1(z),F_2(z)$ 
satisfy the following relations.
\begin{eqnarray}
~[B_{i,m},s_j(z_2)]&=&0,\\
~[E_i(z_1),s_j(z_2)]&=&0,\\
~[F_i(z_1),s_j(z_2)]&=&
\frac{\delta_{i,j}}{(q-q^{-1})z_1z_2}
(\delta(q^{k+1}z_2/z_1)-
\delta(q^{-k-1}z_2/z_1))\nonumber\\
&\times&
:e^{-(\frac{1}{k+1}a^i)(z_1|-\frac{k+1}{2})}
u_i^-(z_1,p)z_1^{\frac{1}{r}(P_i+h_i)}:.
\end{eqnarray}
\end{prop}
From the above proposition we have 
\begin{eqnarray}
\left[\int_0^{s \infty}s_j(z)d_{q^{2(k-1)}}z,
~U_{q,p}(\widehat{sl}(2|1))
\right]=0.
\end{eqnarray}

~\\

\section*{Acknowledgments}

This work is supported by the Grant-in-Aid for
Scientific Research {\bf C} (21540228)
from Japan Society for Promotion of Science. 
The author is grateful to
Pascal Baseilhac 
and the colleagues in University of Tours
for kind invitation and warm hospitality
during his stay in Tours.

\begin{appendix}

\section{Bosonization}

In appendix we summarize relations
of bosonization for $U_q(\widehat{sl}(1|2))$
and its screening currents
relating to the delta-function
$\delta(z)=\sum_{m \in {\mathbb Z}}z^m$.

\begin{eqnarray}
&&
\{x_{1,1}^+(z_1),x_{1,1}^-(z_2)\}\\
&&
=
\frac{q}{z_1}\delta(q^kz_2/z_1)
e^{a_+^{1}(q^{\frac{k-1}{2}}z_2)
-b^{2,3}_+(q^{k-1}z_2)-b^{1,3}_+(q^{k-1}z_2)},
\nonumber
\\
&&
\{x_{1,1}^+(z_1),x_{1,2}^-(z_2)\}\\
&&=
\frac{q}{z_1}\delta(q^{-k+2}z_2/z_1)
:e^{a_-^{1}(q^{-\frac{k-1}{2}}z_2)
-b^{2,3}_+(q^{-k+2}z)-b^{1,3}_+(q^{-k+2}z_2)}:,
\nonumber
\\
&&
[x_{1,1}^+(z_1),x_{2,1}^-(z_2)]
\\
&&=-(q-q^{-1})\delta(q^{k-1}z_2/z_1)
:e^{a_+^2(q^{\frac{k-1}{2}}z_2)
-(b+c)^{2,3}(q^{k-1}z_2)-b_+^{1,2}
(q^{k-1}z_2)-b^{1,2}(q^{k-2}z_2)}:,\nonumber\\
&&\{x_{1,2}^+(z_1),x_{1,3}^-(z_2)\}\\
&&=
\frac{1}{z_1}\delta(q^{-k}z_2/z_1)
e^{a_-^1(q^{-\frac{k-1}{2}}z_2)
-b_-^{2,3}(q^{-k+1}z_2)-b_-^{1,3}(q^{-k+1}z_2)},
\nonumber
\\
&&\{x_{1,2}^+(z_1),x_{1,4}^-(z_2)\}\\
&&=
\frac{1}{z_1}\delta(q^{-k+2}z_2/z_1)
:e^{a_-^1(q^{-\frac{k-1}{2}}z_2)
-b^{2,3}_+(q^{-k+2}z_2)-b^{1,3}_+(q^{-k+2}z_2)}:,
\nonumber\\
&&
[x_{1,2}^+(z_1),x_{2,3}^-(z_2)]\\
&&=-(q-q^{-1})
\delta(q^{k-1}z_2/z_1):
e^{a_+^2(q^{\frac{k-1}{2}}z_2)
-(b+c)^{2,3}(q^{k-1}z_2)-b_+^{1,2}
(q^{k-1}z_2)-b^{1,2}(q^{k-2}z_2)}:,\nonumber
\\
&&[x_{2,1}^+(z_1),x_{1,4}^-(z_2)]\\
&&=-(q-q^{-1})
\delta(q^{-k+1}z_2/z_1)
:e^{a_-^1(q^{-\frac{k-1}{2}}z_2)
+(b+c)^{2,3}(q^{-k}z_2)
+(b+c)^{2,3}(q^{-k+2}z_2)
-b_-^{1,3}(q^{-k+1}z_2)
+b^{1,3}(q^{-k}z_2)}:,\nonumber
\\
&&[x_{2,1}^+(z_1),x_{2,1}^-(z_2)]\\
&&=
(q-q^{-1})
\delta(q^{k}z_2/z_1)
e^{a_+^2(q^{\frac{k-1}{2}}z_2)
+b_+^{2,3}(q^{k-2}z_2)
+b_+^{2,3}(q^{k}z_2)
+b_+^{1,3}(q^{k-2}z_2)
-b_+^{1,2}(q^{k-1}z_2)},
\nonumber\\
&&[x_{2,2}^+(z_1),x_{1,3}^-(z_2)]\\
&&=(q-q^{-1})
\delta(q^{-k+1}z_2/z_1)
:e^{a_-^1(q^{-\frac{k-1}{2}}z_2)
+(b+c)^{2,3}(q^{-k}z_2)
+(b+c)^{2,3}(q^{-k+2}z_2)
-b_-^{1,3}(q^{-k+1}z_2)
+b^{1,3}(q^{-k}z_2)}:,\nonumber
\\
&&[x_{2,2}^+(z_1),x_{1,3}^-(z_2)]\\
&&=-(q-q^{-1})
\delta(q^{-k}z_2/z_1)
e^{a_-^2(q^{-\frac{k-1}{2}}z_2)
+b_-^{2,3}(q^{-k}z_2)
+b_-^{2,3}(q^{-k+2}z_2)
+b_-^{1,3}(q^{-k+2}z_2)
-b_-^{1,2}(q^{-k+1}z_2)}.\nonumber
\end{eqnarray}
\begin{eqnarray}
&&[x_{2,1}^+(z_1),s_{2,3}(z_2)]\\
&&=-(q-q^{-1})
\delta(qz_2/z_1):e^{-(\frac{1}{k-1}a^2)(z_2|\frac{k-1}{2})+
b_-^{1,2}(z)-b_-^{1,3}(qz)+(b+c)^{2,3}(z)+(b+c)^{2,3}(q^2z)}:,
\nonumber\\
&&[x_{2,2}^+(z_1),s_{2,4}(z_2)]\\
&&=(q-q^{-1})
\delta(qz_2/z_1):
e^{-(\frac{1}{k-1}a^2)(z_2|\frac{k-1}{2})+
b_-^{1,2}(z)-b_-^{1,3}(qz)+(b+c)^{2,3}(z)+(b+c)^{2,3}(q^2z)}
:,\nonumber\\
&&[x_{1,1}^+(z_1),s_{2,5}(z_2)]\\
&&=\frac{1}{z_2}\delta(q^2z_2/z_1)
:e^{-(\frac{1}{k-1}a^2)(z_2|\frac{k-1}{2})+b_-^{1,2}(z_2)
-b_+^{1,3}(qz_2)-b^{1,3}(z_2)-b_+^{2,3}(qz_2)}:,\nonumber\\
&&[x_{1,2}^+(z_1),s_{2,3}(z_2)]\\
&&=q^{-1}(q-q^{-1})
\delta(q^2z_2/z_1):
e^{-(\frac{1}{k-1}a^2)(z_2|\frac{k-1}{2})+b_-^{1,2}(z_2)
-b_+^{1,3}(qz_2)-b^{1,3}(z_2)-b_+^{2,3}(qz_2)}:,\\
&&\{x_{1,1}^-(z_1),s_{1,5}(z_2)\}=
\frac{1}{z_2}\delta(q^{-k+1}z_2/z_1)
:e^{-(\frac{1}{k-1}a^1)(z_1|-\frac{k-1}{2})}:,\\
&&
\{x_{1,2}^-(z_1),s_{1,5}(z_2)\}=
\frac{1}{z_2}\delta(q^{k-1}z_2/z_1)
:e^{-(\frac{1}{k-1}a^1)(z_1|-\frac{k-1}{2})}:,\\
&&
[x_{2,1}^-(z_1),s_{2,3}(z_2)]=
(q-q^{-1})\delta(q^{-k+3}z_2/z_1)
\nonumber\\
&&\times
:e^{a_+^2(q^{\frac{k-1}{2}}z_1)
-(\frac{1}{k-1}a^2)(q^{k-3}z_1|\frac{k-1}{2})
+b_+^{1,3}(q^{k-2}z_1)-b_-^{1,3}(q^{k-2}z_1)
-b_+^{1,2}(q^{k-1}z_1)+b_-^{1,2}(q^{k-3}z_1)}:),
\\
&&
[x_{2,2}^-(z_1),s_{2,4}(z_2)]=
-(q-q^{-1})\delta(q^{k-1}z_2/z_1)
:e^{-(\frac{1}{k-1}a^2)(z_1|-\frac{k-1}{2})}:,\\
&&
[x_{2,3}^-(z_1),s_{2,5}(z_2)]=
\frac{-1}{(q-q^{-1})q^{k-2}z_1z_2}\nonumber\\
&&\times(
\delta(q^{-k+1}z_2/z_1)
:e^{-(\frac{1}{k-1}a^2)(z_1|-\frac{k-1}{2})}:
\\
&&-
\delta(q^{-k+3}z_2/z_1)
:e^{a_+^2(q^{\frac{k-1}{2}}z_1)
-(\frac{1}{k-1}a^2)(q^{k-3}z_1|\frac{k-1}{2})
+b_+^{1,3}(q^{k-2}z_1)-b_-^{1,3}(q^{k-2}z_1)
-b_+^{1,2}(q^{k-1}z_1)+b_-^{1,2}(q^{k-3}z_1)}:),
\nonumber\\
&&
[x_{1,2}^-(z_1),s_{2,3}(z_2)]\\
&&=(q-q^{-1})
\delta(q^kz_2/z_1):
e^{a_-^1(q^{\frac{k+1}{2}}z_2)
-(\frac{1}{k-1}a^2)(z_2|\frac{k-1}{2})
+b_-^{1,2}(z_2)+b^{1,2}(qz_2)-b_-^{1,3}(qz_2)-b_+^{2,3}(qz_2)
+(b+c)^{2,3}(q^2z_2)}:,\nonumber\\
&&
[x_{1,2}^-(z_1),s_{2,4}(z_2)]\\
&&=(q-q^{-1})
\delta(q^kz_2/z_1):
e^{a_-^1(q^{\frac{k+1}{2}}z_2)
-(\frac{1}{k-1}a^2)(z_2|\frac{k-1}{2})
+b_-^{1,2}(z_2)+b^{1,2}(qz_2)-b_-^{1,3}(qz_2)-b_-^{2,3}(qz_2)
+(b+c)^{2,3}(q^2z_2)}:,\nonumber
\\
&&
[x_{1,3}^-(z_1),s_{2,3}(z_2)]=-q(q-q^{-1})
\delta(q^kz_2/z_1)\\
&&\times:
e^{a_-^1(q^{\frac{k+1}{2}}z_2)
-(\frac{1}{k-1}a^2)(z_2|\frac{k-1}{2})
+b_-^{1,2}(z_2)-2b_-^{1,3}(qz_2)-b_-^{2,3}(qz_2)-b_+^{2,3}(qz_2)
+(b+c)^{2,3}(z_2)+(b+c)^{2,3}(q^2z_2)}:,\nonumber\\
&&
[x_{1,3}^-(z_1),s_{2,5}(z_2)]\\
&&=\frac{1}{z_2}
\delta(q^kz_2/z_1):
e^{a_-^1(q^{\frac{k+1}{2}}z_2)
-(\frac{1}{k-1}a^2)(z_2|\frac{k-1}{2})
+b_-^{1,2}(z_2)+b^{1,2}(qz_2)-b_-^{1,3}(qz_2)-b_-^{2,3}(qz_2)
+(b+c)^{2,3}(q^2z_2)}:,\nonumber
\\
&&
[x_{1,4}^-(z_1),s_{2,4}(z_2)]=q(q-q^{-1})\delta(q^kz_2/z_1)
\\
&&\times
:e^{a_-^1(q^{\frac{k+1}{2}}z_2)
-(\frac{1}{k-1}a^2)(z_2|\frac{k-1}{2})
+b_-^{1,2}(z_2)-2b_-^{1,3}(qz_2)-b_-^{2,3}(qz_2)-b_+^{2,3}(qz_2)
+(b+c)^{2,3}(z_2)+(b+c)^{2,3}(q^2z_2)}:,\nonumber\\
&&
[x_{1,4}^-(z_1),s_{2,5}(z_2)]\\
&&=\frac{1}{z_2}\delta(q^kz_2/z_1)
:e^{a_-^1(q^{\frac{k+1}{2}}z_2)
-(\frac{1}{k-1}a^2)(z_2|\frac{k-1}{2})
+b_-^{1,2}(z_2)+b^{1,2}(qz_2)-b_-^{1,3}(qz_2)-b_+^{2,3}(qz_2)
+(b+c)^{2,3}(q^2z_2)}:.\nonumber
\end{eqnarray}

~\\

\end{appendix}


\begin{thebibliography}{99}
\bibitem{BPZ}A.A.Belavin, A.M.Polyakov and A.B.Zamolodchikov,
Infinite Conformal Symmetry 
in Two-dimensional Quantum Field Theory,
{\it Nucl.Phys.} {\bf B241} 333-380 (1984).
\bibitem{JM}M.Jimbo and T.Miwa,
{\it Algebraic Analysis of Solvable Lattice Models},
CBMS Regional Conference Series
in Mathematics {\bf 85} 
(American Mathematical Society), 1994.
\bibitem{ABF}G.E.Andrews, R.J.Baxter and P.J.Forrester,
Eight-Vertex SOS Model and Generalized Rogers-Ramanujan-type
identities,
{\it J.Stat.Phys.} {\bf 35}, 193-266, (1984).
\bibitem{Konno}H.Konno,
An Elliptic Algebra $U_{q,p}(\widehat{sl}_2)$
and the Fusion RSOS model,
{\it Commun.Math.Phys.} {\bf 195},
373-403, (1998).
\bibitem{JKOS2}M.Jimbo, H.Konno, S.Odake and J.Shiraishi,
Elliptic Algebra $U_{q,p}(\widehat{sl_2})$ :
Drinfeld Currents and Vertex Operators,
{\it Commun.Math.Phys.} {\bf 199}, 605-647, (1999).
\bibitem{Baxter}R.Baxter,
Exactly Solved Models in Statistical Mechanics,
Academic Press, London, 1982. 
\bibitem{DJMO}E.Date, M.Jimbo, T.Miwa and M.Okado,
Fusion of the Eight Vertex SOS Model,
{\it Lett.Math.Phys.}{\bf 12}, 209-215, (1986).
\bibitem{DJKMO}E.Date, M.Jimbo, A.Kuniba, T.Miwa and M.Okado,
Exactly Solvable SOS Models. II: Proof of the 
Star-triangle Relation
and Combinatorial Identities,
{\it Adv.Stu.Pure.Math.}
{\bf 16}, 17-122, (1988).
\bibitem{DJKMO2}
E.Date, M.Jimbo, A.Kuniba, T.Miwa and M.Okado,
Exactly Solvable SOS Models :
Local Height Probabilities and Theta Function Identities,
{\it Nucl.Phys.}{\bf B290}[FS20],
231-273, (1987).
\bibitem{JMO}M.Jimbo, T.Miwa and M.Okado,
Solvable Lattice Models whose States are
Dominant Integral Weights of $A_{n-1}^{(1)}$,
{\it Lett.Math.Phys.}{\bf 14},
123-131, (1987).
\bibitem{JKMO}M.Jimbo, A.Kuniba, T.Miwa and M.Okado,
The $A_n^{(1)}$ Face Models,
{\it Commun.Math.Phys.}
{\bf 119}, 543-565, (1988).
\bibitem{DFJMN}B.Davies, O.Foda, M.Jimbo, T.Miwa 
and A.Nakayashiki,
Diagonalization of the XXZ Hamiltonian by Vertex operators,
{\it Commun.Math.Phys.}{\bf 151}, 89-153, (1993).
\bibitem{JMMN}M.Jimbo, K.Miki, T.Miwa and A.Nakayashiki,
Correlation Functions of the XXZ Model for $\Delta<-1$,
{\it Phys.Lett.}{\bf B168}, 256-263, (1992).
\bibitem{FJMMN}O.Foda, M.Jimbo, T.Miwa, K.Miki 
and A.Nakayashiki,
Vertex Operators in Solvable Lattice Models,
{\it J.Math.Phys.}{\bf 35}, 13-46, (1994).
\bibitem{LP}S.Lukyanov and Ya.Pugai,
Multi-point Local Height Probabilities in the Integrable 
RSOS Model,
{\it Nucl.Phys.}{\bf B473}, 631-658, (1996).
\bibitem{AJMP}Y.Asai, M.Jimbo, T.Miwa and Ya.Pugai,
Bosonizations of Vertex Operators for $A_{n-1}^{(1)}$
Face Model,{\it J.Phys.}{\bf A29}:Math.Gen.,
6595-6616, (1996).
\bibitem{FKQ}H.Furutsu, T.Kojima and Y.-H.Quano,
Type-II Vertex Operators for
the $A_{n-1}^{(1)}$ Face Model,
{\it Int.J.Mod.Phys.}{\bf A15}, 1533-1556, (2000).
\bibitem{FJMOP}B.Feigin, M.Jimbo, T.Miwa, A.Odesskii
and Ya.Pugai,
Algebra of Screening Operators for the Deformed $W_n$ Algebra,
{\it Commun.Math.Phys.}{|bf 191}, 501-541, (1998).
\bibitem{Okado}M.Okado,
Solvable Face Models Related to the Lie Superalgebra
$sl(m|n)$,
{\it Lett.Math.Phys.}
{\bf 22}, 39-43, (1991).
\bibitem{D}V.G. Drinfeld,
A New Realization of Yangians and Quantized Affine
Algebras,
{\it Sov.Math.Dokl.} {\bf 36},
212-216, (1988).
\bibitem{KK1}T.Kojima and H.Konno,
The Elliptic Algebra $U_{q,p}(\widehat{sl}_N)$
and the Drinfeld Realization of Elliptic Quantum Group
${\cal B}_{q,\lambda}(\widehat{sl}_N)$,
{\it Commun.Math.Phys.} {\bf 239}, 405-447, (2003).
\bibitem{KK2}T.Kojima and H.Konno,
The Drinfeld Realization of the Elliptic Quantum Group
${\cal B}_{q,\lambda}(A_2^{(2)})$,
{\it J.Math.Phys.} {\bf 45}, 3146-3179 ,(2004).
\bibitem{Sklyanin}E.K.Sklyanin,
Some Algebraic Structure Connected with 
the Yang-Baxter Equation,
{\it Funct.Anal.Appl.} {\bf 16}, 263-270, (1982).
\bibitem{FIJKMY}O.Foda, K.Iohara, M.Jimbo, R.Kedem,
T.Miwa and H.Yan,
An Elliptic Quantum Algebra for
$\widehat{sl}_2$,
{\it Lett.Math.Phys.} {\bf 32}, 259-268, (1994).
\bibitem{Felder}G.Felder,
Elliptic Quantum Group, {\it Proc.ICMP Paris 1994},
Cambridge-Hong Kong :
International Press,
211-218, (1995).
\bibitem{EF}B.Enriquez and G.Felder,
Elliptic Quantum Groups $E_{\tau,\eta}(\widehat{sl}_2)$
and Quasi-Hopf Algebras,
{\it Commun.Math.Phys.} {\bf 195}, 
651-689, (1998).
\bibitem{Fronsdal1}C.Fr$\phi$nsdal,
Generalization and Exact Deformations of Quantum Groups,
{\it Publ.Res.Math.Sci.} {\bf 33},
91-149, (1997).
\bibitem{Fronsdal2}C.Fr$\phi$nsdal,
Quasi-Hopf Deformation of Quantum Groups,
{\it Lett.Math.Phys.} {\bf 40},
117-134, (1997).
\bibitem{JKOS1}M.Jimbo, H.Konno, S.Odake and J.Shiraishi,
Quasi-Hopf Twistors for Elliptic Quantum Groups,
{\it Transformation Groups} {\bf 4}, 303-327, (1999).
\bibitem{ZG}Y.-Z. Zhang and M.D.Gould,
Quasi-Hopf Superalgebras and Elliptic Quantum Superalgebra,
{\it J.Math.Phys.} {\bf 40}, 5264-5282, (1999).
\bibitem{Konno2}H.Konno,
Elliptic quantum group $U_{q,p}(\widehat{sl}_2)$,
Hopf algebroid structure and elliptic hyper geometric series,
{\it J. Geometry and Physics} {\bf 59},
1485-1511, (2009).
\bibitem{Konno3}H.Konno,
Elliptic quantum group and vertex operators,
{\it J.Phys.}{\bf A41}:Math.Theor.
194012(12 pages), (2008).
\bibitem{Miki}K.Miki,
Creation/annihilation Operators and Form Factors of the XXZ
model, {\it Phys.Lett.}{\bf A186}, 217-224, (1994).
\bibitem{Kac-Peterson}
V.G.Kac and D.H.Peterson,
Infinite-dimensional Lie Algebras, Theta Functions and Modular Forms,
{\it Adv.Math.}{\bf 53}, 125-264, (1984).
\bibitem{GKO}P.Goddard, A.Kent and D.Olive,
Virasoro Algebras and Coset Space Models,
{\it Phys.Lett.}{\bf B152}, 88-92, (1985).
\bibitem{Serganova}
V.Serganova,
Kazhdan-Lusztig polynomials and character formula
for the Lie superalgebra $gl(m|n)$,
{\it Selecta Math.}{\bf 2}, 607-651, (1996).
\bibitem{Jonathan}
B.Jonathan,
Kazhdan-Lusztig polynomials and character formula
for the Lie superalgebra $gl(m|n)$,
{\it J.Amer.Math.Sci.}{\bf 16}, 185-231, (2003).
\bibitem{Kac}V.G.Kac,
Lie Superalgebras,
{\it Advances in Math.} {\bf 26}, 8-96, (1977).
\bibitem{Leur}Johan W.van de Leur,
A Classification of Contragredient Lie Superalgebras of Finite Growth,
{\it Commun. in Algebra} {\bf 17},
1815-1841, (1989).
\bibitem{Yamane}H.Yamane,
On Defining Relations of the affine Lie Superalgebras
and their Quantized Universal Enveloping Superalgebras,
{\it Publ.Res.Inst.Math.Sci.} {\bf 35}, 321-390, (1999).
\bibitem{Wakimoto}M.Wakimoto,
Fock Representation of the Affine Lie Algebra $A_1^{(1)}$,
{\it Commun.Math.Phys.} {\bf 104},
605-609, (1986).
\bibitem{FF}B.L.Feigin and E.V.Frenkel,
Representation of Affine Kac-Moody Algebras and Bosonization,
Physics and Mathematics of Strings (World Scientific),
271-316, 1990.
\bibitem{FF2}
B.L.Feigin and E.V.Frenkel, 
Affine Kac-Moody Algebras and Semi-Infinite Flag Manifolds,
{\it Commun.Math.Phys.} {\bf 128}, 161-189, (1990).
\bibitem{Shiraishi}J.Shiraishi,
Free Boson Representation of $U_q(\widehat{sl}_2)$,
{\it Phys.Lett.} {\bf A171}, 243-248, (1992).
\bibitem{AOS1}H.Awata, S.Odake and J.Shiraishi,
Free Boson Realization of $U_q(\widehat{sl}_N)$,
{\it Commun.Math.Phys.} {\bf 162}, 61-83, (1994).
\bibitem{AOS2}H.Awata, S.Odake and J.Shiraishi,
$q$-Difference Realization of $U_q(sl(M|N))$
and Its Application
to Free Boson Realization of $U_q(\widehat{sl}(2|1))$,
{\it Lett.Math.Phys.} {\bf 42}, 271-279, (1997).
\bibitem{ZG2}Y.-Z.Zhang and M.D.Gould,
$U_q(\widehat{sl}(2|1))$ Vertex Operators, Screen Currents 
and Correlation Functions at Arbitrary Level,
{\it J.Math.Phys.}{\bf 41}, 5277-5291, (2000).


\end{thebibliography}
\end{document}